\documentclass[onecolumn]{aastex631}
\usepackage{amsmath}
\usepackage{mathtools}

\def\Ha{{\rm H}\alpha}
\def\Hb{{\rm H}\beta}
\def\secref#1{\S\ref{#1}}

\shorttitle{Diversity of Galaxy Spectra}
\shortauthors{Teimoorinia et al.}

\graphicspath{{./}{figures/}}
\usepackage{wrapfig}

\begin{document}

\title{Mapping the Diversity of Galaxy Spectra with Deep Unsupervised Machine Learning}

\correspondingauthor{Hossen Teimoorinia}
\email{hossen.teimoorinia@nrc-cnrc.gc.ca, hossteim@uvic.ca}

\author{Teimoorinia, Hossen}
\affiliation{NRC Herzberg Astronomy and Astrophysics, 5071 West Saanich Road, Victoria, BC, V9E 2E7, Canada }
\affiliation{Department of Physics and Astronomy, University of Victoria, Victoria, BC, V8P 5C2, Canada}

\author{Archinuk, Finn}
\affiliation{NRC Herzberg Astronomy and Astrophysics, 5071 West Saanich Road, Victoria, BC, V9E 2E7, Canada }

\author{Woo, Joanna}
\affiliation{Department of Physics, Simon Fraser University, 8888 University Dr, Burnaby BC, V5A 1S6, Canada}

\author{Shishehchi, Sara}
\affiliation{Pegasystems Inc. 1 Main Street, Cambridge, MA 02142, USA}
 
\author{Bluck, Asa F. L.}
\affiliation{Kavli Institute for Cosmology \& Cavendish Laboratory Astrophysics Group, University of Cambridge, Madingley Road, Cambridge, CB3 OHA, UK}

\begin{abstract}
Modern spectroscopic surveys of galaxies such as MaNGA consist of millions of diverse spectra covering different regions of thousands of galaxies. 
We propose and implement a deep unsupervised machine learning method to summarize the entire diversity of MaNGA spectra onto a 15x15 map (DESOM-1), where neighbouring points on the map represent similar spectra.  We demonstrate our method as an alternative to conventional full spectral fitting for deriving physical quantities, as well as their full probability distributions, much more efficiently than traditional resource-intensive Bayesian methods.  Since spectra are grouped by similarity, the distribution of spectra onto the map for a single galaxy, i.e, its ``fingerprint", reveals the presence of distinct stellar populations within the galaxy indicating smoother or episodic star-formation histories.  We further map the diversity of galaxy fingerprints onto a second map (DESOM-2).  Using galaxy images and independent measures of galaxy morphology, we confirm that galaxies with similar fingerprints have similar morphologies and inclination angles.  Since morphological information was not used in the mapping algorithm, relating galaxy morphology to the star-formation histories encoded in the fingerprints is one example of how the DESOM maps can be used to make scientific inferences.
\end{abstract}


\section{Introduction}
\label{sec:introduction}

Galaxy spectra are undoubtedly crucial for understanding the properties and evolution of galaxies. Apart from redshift, galaxy spectra contain a vast amount of information about the galaxy's history, including the metallicity content \citep[e.g., ][]{Tremonti04, Gallazzi06, Kewley08, teimoorinia21a}, star formation history \citep[e.g., ][]{Kauffmann03, Salim05, Conroy2013}, stellar and gas kinematics, and gas ionization by star formation and active galactic nuclei (AGN) \citep[e.g., ][]{Kewley08}.

Traditionally, a single spectrum is observed for each galaxy, either by long-slit or fibre spectroscopy (e.g., for the Sloan Digital Sky Survey - \citealp{York2000,Gunn2006}).  The advent of large integral field unit (IFU) surveys of nearby galaxies, such as Mapping Nearby Galaxies at the Apache Point Observatory (MaNGA - \citealp{Drory15,Bundy2015}), Calar Alto Legacy Integral Field spectroscopy Area (CALIFA - \citealp{Sanchez2012}) and Sydney-AAO Multi-object Integral field spectrograph survey (SAMI - \citealp{Croom2012}), has allowed a more detailed study of spectrum-derived properties for individual regions within galaxies.  

However, the increasingly detailed view of galaxies made possible by IFU surveys has required vastly more computational resources to process.  In order to properly sample the line spread function of a mid-resolution spectrograph (such as the ones employed for MaNGA) over the entire optical wavelength range, a single spectrum will consist of over 4000 data points for the flux measurements.  For the MaNGA dataset, each galaxy is usually covered by hundreds to thousands of such spectra, depending on the angular size of the galaxy. The public Data Release 15 of the MaNGA survey currently includes about 4600 galaxies, and 10 000 in total are planned \citep{Drory15,Bundy2015}.  Full spectral analysis of such data sets using conventional tools strains current computational resources.

Machine learning (ML) approaches to data analysis are not only orders of magnitude faster than conventional methods, but can also solve complex and challenging problems.  There has been an explosion of interest in ML in many fields of astronomy. For example, ML provides an effective tool to rank tabular data sets \citep[e.g.,][]{Teimoorinia16, Bluck20}, and classification and clustering of spectral data \citep[e.g.,][]{teimoorinia12,rahmani18}. Deep learning methods have also proven that they can explore complicated images to find new and rare phenomena  \citep[e.g.,][]{Pourrahmani18,Jacobs19,Bottrell19, Teimoorinia20b,Bickley21} or to classify large sets of astronomical images in a fully deep unsupervised mode \citep[e.g.,][]{Teimoorinia21b}. 

A common application of ML methods is the classification problem (e.g., ``what type of spectrum is this?"), which can be solved using supervised and unsupervised methods.
In the supervised case, ML requires a training set of spectra with labels (e.g., ``AGN", ``star-forming", ``passive", ``poststarbrust") given by human classifiers or by other conventional methods. A clear disadvantage to such supervised methods is that labels must be provided using methods that are usually subject to a host of biases and incompleteness. Finer class divisions that are currently unknown to astronomers could potentially hide important astrophysical understanding.

Unsupervised methods find patterns within the raw data with minimum assumptions in order to determine how to divide the data.  Post-facto, astronomers may find that the spectral classes found by the unsupervised method correspond to ``AGN" or ``star-forming" spectra, but providing these categories are unnecessary.  The astronomer merely determines beforehand how many classes are desired.  One such unsupervised clustering method is the Self-Organising Map (SOM).  The machine-determined classes are organized onto a map where neighbouring regions of the map have more similar classes of spectra than non-neighbouring regions.  The human astronomer decides the dimensionality of the map.  Most SOM practitioners use a two-dimensional map, as we use here, but the method could in principle be extended to three (or more) dimensions.  SOMs are often used as a data visualization method, but as we will show here, when applied to astronomical spectra, the nodes on the map also contain  scientific information about the distribution of spectra.

How do these machine-determined spectral classes correspond to known phenomena such as AGN, or to more continuous labels such as the star formation rate (SFR), stellar age, or metallicity?  Independent measurements of these indicators/quantities can be determined by traditional methods such as full spectral fitting (e.g., using the Penalized Pixel Fitting or pPXF code - \citealp{Cappellari2017}) for individual spectra and averaged within different regions of the SOM.  Then, given a spectrum with unknown stellar age, for example, one need only locate the spectrum on the SOM in order to determine the likelihood distribution of the stellar age.  Although this method of deriving physical quantities from spectra is subject to the same biases inherent in the traditional methods (e.g., pPXF), it is much faster than said traditional methods.  Labelling is a separate step and different labelling methods (e.g., \textsc{Starlight} - \citealp{CidFernandes2005} - instead of pPXF) can be applied on top of this map without needing to retrain. 

In this paper we demonstrate an unsupervised method for organizing the spectra of the MaNGA survey (DR15), and show its potential for answering astrophysical questions.  MaNGA is currently the largest IFU survey to date, and has produced hundreds of papers since the first observations in 2015.  Since traditional SOMs cannot handle high-dimensional data (i.e., the $> 4000$ wavelength points for a single spectrum), we use a Deep Embedded Self-Organizing Map \citep[DESOM;][]{Forest2019}, that first reduces the dimensionality of  data before dividing them into a SOM.  The DESOM not only successfully groups spectra, but effectively shows the correlations and the link between different groups, based only on features in the spectra.  We will show how the DESOM can summarize the entirety of the MaNGA-observed universe into a 15$\times$15 grid of spectra. For example, one application of this map is to provide high S/N template spectra, which we will show in \secref{sec:appfitting}. We map pPXF-measured quantities onto the DESOM map to demonstrate that well-separated areas of the map correspond to very different values in pPXF-measured quantities.  Each node in the grid effectively constitutes a unique spectral template to which an observed spectrum with unknown labels can be compared to in order to determine its likely labels.

The defining characteristic of MaNGA as a spectroscopic survey is of course its spatially resolved nature.  
We extend the DESOM method to take advantage of this spatial information by noting that the spectra within a single galaxy fall onto our DESOM map in a unique pattern which we call the ``fingerprint" for this galaxy.  Since the spectra on the DESOM are grouped by similarity, one may immediately discern, for example, the existence of multiple distinct stellar populations, indicating an episodic star formation history.
We then demonstrate how the DESOM fingerprints of galaxies can themselves be organized onto a second DESOM.  The map produced by the second DESOM completely describes the MaNGA-observed diversity of galaxies according to their spatially resolved spectra.

This paper is organized as follow: we introduce the data used in this paper in \secref{sec:data}. We describe the deep learning method in \secref{sec:method}. 
We then present three applications of this model in \secref{sec:results}.

\section{Data}
\label{sec:data}

Our sample is drawn from Data Release 15 (DR15) of the MaNGA IFU survey \citep{Bundy2015}.  This data release contains 4609 unique galaxies that have a redshift in the NASA Sloan Atlas, covered by over 9 million spaxels with varying levels of continuum signal-to-noise ratio (S/N).  We used the Voronoi binning code of \cite{Cappellari2003a} to adaptively bin adjacent spaxels in a single galaxy so that the resulting bin of spaxels (called ``baxels", following \citealp{Woo2019}) achieved a target S/N of 20.  Before binning, we masked foreground stars as well as individual spaxels having S/N $<$ 2.  This latter masking improved the contiguity of the binning maps of some galaxies with many spaxels of low S/N, and was recommended by the documentation of Voronoi binning code.  

To account for the covariance of the noise between adjacent spaxels due to the MaNGA's sampling method, \cite{Law2016} computed a correction factor which we multiplied to the quadrature-added noise of a resulting baxel.  When the number $n$ of spaxels that are contained in a baxel is less than 100, this correction factor is $1.0 + 1.62\log(n)$. For $n > 100$, the correction factor is a constant 4.2.

The binning process resulted in 1 170 539 baxels across 4609 galaxies.
The number of baxels per galaxy varies from galaxy to galaxy, with the large majority having fewer than 500 baxels.  56 galaxies were observed more than once; most of these were observed twice, but some were observed 3 or 4 times.  Thus the entire data set contains 4682 data cubes.  We did not attempt to combine cubes of the same galaxy and each cube was binned separately.  Therefore some areas of a single galaxy may be covered by more than one baxel.  This double counting represents 3\% of baxels, but does not affect our demonstration of the DESOM method.

MaNGA galaxies cover a redshift range of 0 to 0.15.  In order to sample a uniform rest-frame wavelength range for all galaxies, we trimmed the spectra to a rest-frame wavelength range of 3650 to 8950 \AA, which safely falls within the observed wavelength range for all galaxies.  Since convolutional neural networks of the type we implement here can be trained to recognize spectral features regardless of redshift, this de-redshifting of the spectra is not strictly necessary.  However, our purpose here is to demonstrate the power of the simplest DESOM, and therefore defer a demonstration of redshift invariance to a future study. 

ML implementations are greatly simplified when the data has uniform dimensions, preferably in a multiple of a high power of 2. We rebinned the baxel spectra to 4544 wavelength bins (within the restframe 3650 to 8950 \AA) which is a multiple of $2^6$.  Before rebinning, we masked the regions of the prominent terrestrial oxygen lines at 5577 and 6300 \AA, as well as NaD emission at 5895 \AA.

\begin{figure*}
\centering
\includegraphics[height=20cm,angle=0]{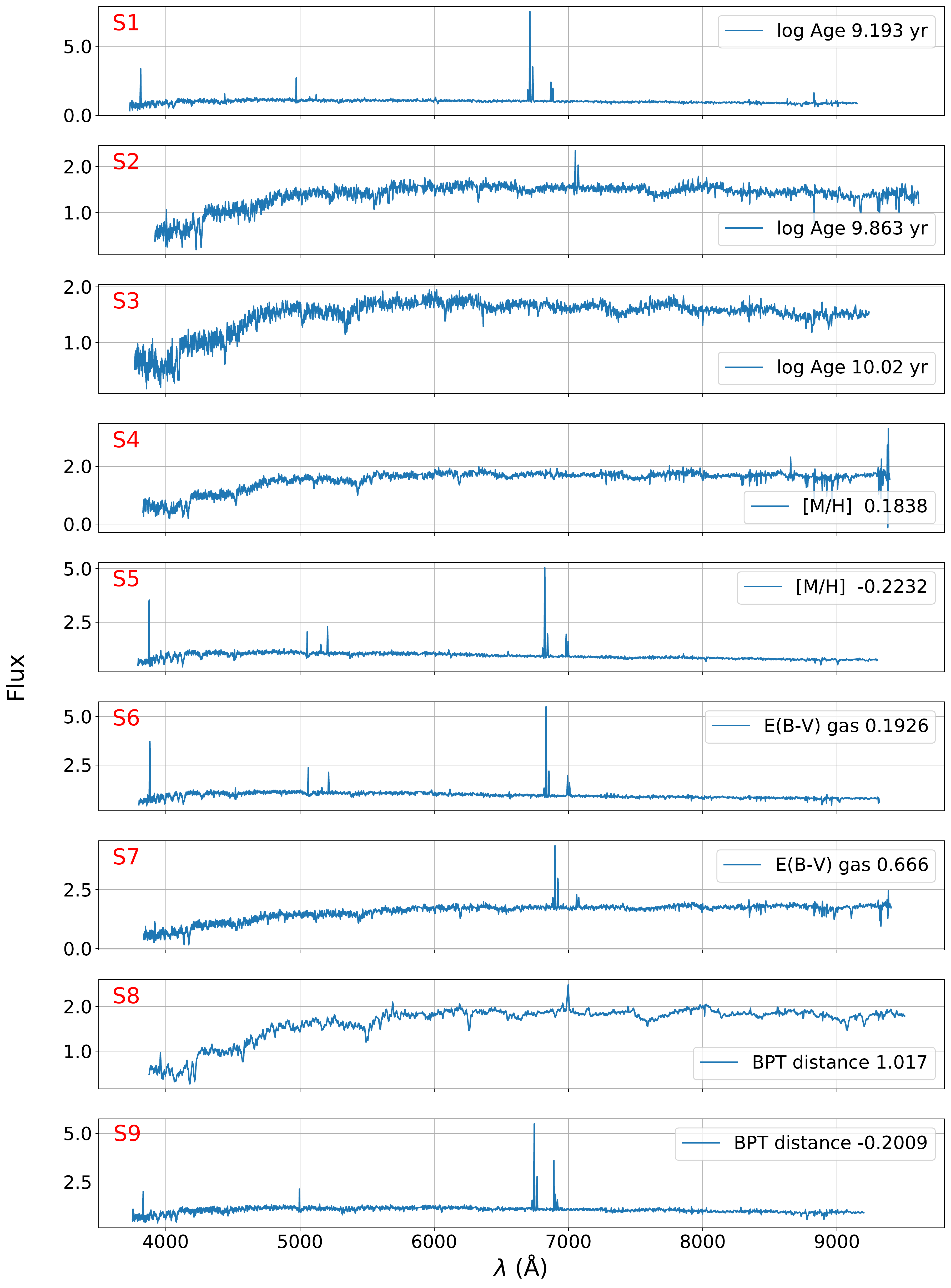}
\vspace{.5cm}
\caption{A selection of spectra from the MaNGA dataset. These were selected to show the range of a few physical parameters as estimated by full spectrum fitting by pPXF.}
\label{fig:9spectra}
\end{figure*}

Baxel spectra were normalized by the mean flux between 4050 and 4150\AA~ so that the continuum of each spectrum would be roughly the same level, which aids in network training.  The emission line levels can vary widely.  Figure \ref{fig:9spectra} shows a sample of normalized baxel spectra. 

We removed 2\% of the baxels from our final sample for the following reasons.  Some spectra had bad flux values in too many wavelength bins, as flagged by the MaNGA data reduction pipeline.  We removed all baxels that had fewer than 3900 non-flagged flux values (1.8\%).  
We removed baxels that had strongly negative flux values due to sky residuals in the red part of the spectrum.  We chose to remove baxels with negative fluxes in the lowest 0.5 percentile. 
We also removed baxels with extremely high flux relative to the continuum due to pipeline abnormalities (also in the 0.5 percentile), but only removed the baxel if the high flux values did not correspond to known galaxy emission lines. This high flux cut removed 0.3\%.

Lastly, although our target S/N of the Voronoi binning was 20, the target was not reached for many baxels, mostly in the outskirts of galaxies.  Therefore we removed baxels with S/N $<$10 for our training set which were 5.6\% of spectra.  We tried training models with less restrictive cuts (SNR $<$5) but found that our qualitative results are not significantly changed.

In order to check the smoothness of our DESOM map, we performed full spectral fitting on each baxel using the pPXF code of \cite{Cappellari2017} (v6.7.13\footnote{We used the same small modification used in \cite{Woo2019} to remove the limits on the coefficients of the multiplicative polynomial, in order to allow for cases of extreme reddening.}) following the procedure of \cite{Woo2019}.  We use the attenuation curve of \cite{Fitzpatrick1999} to correct for foreground extinction.  We fit the baxels using the E-MILES templates of \cite{Vazdekis2016} with the \cite{Kroupa2001} IMF and BASTI theoretical isochrones \cite{Pietrinferni2004}.  We refer the reader to \cite{Woo2019} for further details of the full spectral fitting.  

pPXF computes the linear combination of stellar population templates that produces the best-fitting spectrum, while simultaneously fitting the emission lines.  The weights of the linear combination of templates can be used to compute the mass-weighted mean of several stellar quantities: age, total metallicity ([M/H]) and mass-to-light ratio in the r-band ($M_*/L_r$).  In addition, we estimate the redenning of starlight E(B-V)$_{*}$ from the shape of the continuum (from the best-fit multiplicative polynomial from pPXF) using the \cite{Calzetti2000} reddening law, and assuming that the flux calibration of the spectra is reasonable. 

Using the best fitting emission lines, we computed the nebular reddening E(B-V)$_{\rm gas}$ using the Balmer decrement and the \cite{Cardelli1989} extinction curve which is more appropriate for gas \citep{Steidel2014}.  We then computed the specific star-formation rate (sSFR) from the dust-corrected $\Ha$ flux.  Lastly, the diagnostic diagram of \cite{Baldwin1981} (BPT), consisting of the line ratios [NII]6583/$\Ha$ vs. [OIII]5007/$\Hb$, is often used to distinguish between ionization sources, especially AGN vs young stars.  For each baxel, we computed the perpendicular distance ($d_{\rm BPT}$) of their line ratios in this diagram from the curve defined by \cite{Kauffmann2003a}.  Star-forming baxels have $d_{\rm BPT} \lesssim 0$, AGN have $d_{\rm BPT} \gtrsim 0.3$, while intermediate values indicate composite spectra.

\section{The DESOM Algorithm}
\label{sec:method}

Our first goal is to organize the baxels onto a map sorted only by features in their spectra. Since our data contain $>$ 4500 wavelength points on a single spectrum, and SOMs are more efficient on low dimensional data, we must reduce the complexity of the data before training the SOM.  For this reason, we use the Deep Embedded Self Organizing Map (DESOM) method, which is explained in detail in \cite{Forest2019}, but which we also briefly describe here.  

The DESOM method uses an autoencoder to reduce the dimensionality of MaNGA spectra.  An autoencoder is a type of unsupervised neural network whose goal is to represent the input data at a desired lower dimension, as determined by the user.  The quality of an autoencoder is determined by its ability to reconstruct the spectrum as measured by the mean-squared error (MSE) between the input and output. Wider latent widths decrease the error, but at the cost of our main goal of generating a compact representation.  \cite{Portillo} compare reconstruction quality of spectra using autoencoders to PCA and find that the autoencoders only perform better when the latent width is $\lesssim$10. However, the architectures they propose do not utilize convolutional layers, which help summarize spatial features. By incorporating convolutional layers, we find the reconstruction quality of our autoencoder is $\sim 20$\% better than PCA using a latent width of 256 components.
A latent width of 256 (i.e., a 256-dimensional vector) best balances the conflicting demands of dimensionality reduction and reconstruction quality. The encoder part of the autoencoder learns to approximate the otherwise unknown function that converts the spectrum into this 256-dimensional representation. In order to check that the representation truly does describe the original spectrum, a decoder attempts to reconstruct the original spectrum from this representation. A comparison of the output of the decoder to the original spectrum trains the autoencoder to find the encoding function. We use the MSE to evaluate the quality of the reconstruction from the decoder.

\begin{figure*}
\centering
\includegraphics[width=16cm,angle=0]{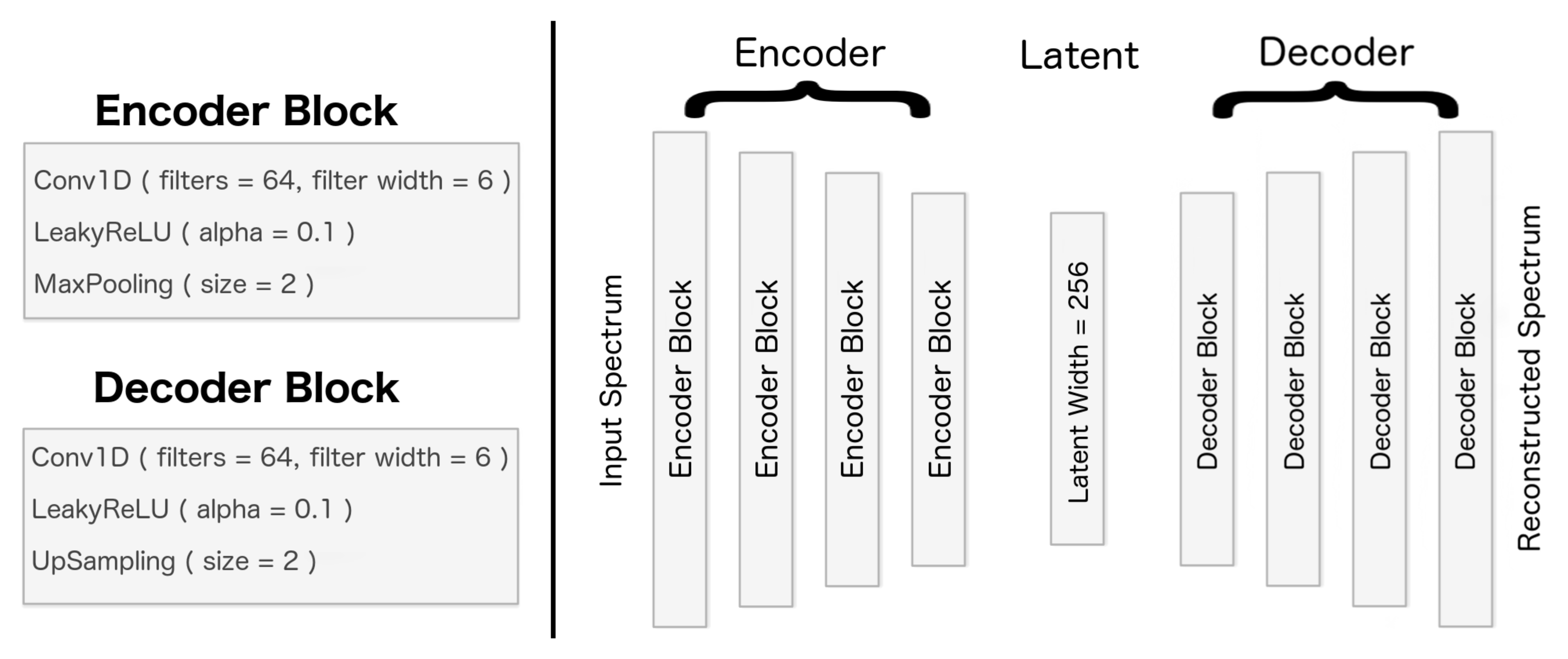}
\caption{The architecture of our autoencoder component uses convolutional blocks. The encoder embeds data to the latent layer. The decoder takes the latent layer and attempts to reconstruct the original input.}
\label{fig:autoencoder}
\end{figure*}

Figure \ref{fig:autoencoder} shows the architecture of the autoencoder that we use in our DESOM.  The encoder consists of four convolutional blocks (or layers) that reduce the dimensionality to the latent width, followed by four convolutional blocks that increase this back to the original dimensions. By stacking these blocks, we can build up more abstract representations, which will be required to compress the input into only 256 dimensions. Convolutional layers allow for translational invariance, which is important for absorption and emission lines since small misalignment errors can still be understood by the DESOM.

The architecture of our autoencoder also includes a ``max pooling" layer, which selects the maximum value within a given window size and thus results in a loss of information.  The purpose of such an operation is to select the most important features of the input data while reducing its dimensionality.

An important parameter for an autoencoder is the width of its latent layer, which we have chosen to be 256.  A wider latent width allows the autoencoder to more accurately reconstruct the output, but accuracy needs to be balanced with the dimensional limitations of the SOM.  By measuring the quality of the decoder's reconstructed spectra as a function of latent width, we can decide how much compression is acceptable.  Our choice of the latent width of 256 is informed by the observation that our reconstruction quality increases with latent width, but begins to plateau around this point.   

Having reduced each spectrum from 4544 dimensions to 256, this compressed representation can now be passed to a SOM.  Since the representation of the spectra impacts the SOM, \cite{Forest2019} argues for training both the autoencoder and SOM simultaneously which we demonstrate below.
The dimensions of the output map are a hyperparameter defined outside of training by the user.  Since SOMs are useful for discretizing continuous data, it can be difficult to know how many nodes are necessary.  We have chosen a 15x15 map as output to ensure enough spectra (at least $\sim 400$) fall into each node to allow statistically robust labelling. We could have chosen a larger map, but we will argue in \secref{sec:appd2} that a smaller SOM is preferable.
Each node on the output map can be represented by a vector with the same dimensions as the latent data. In SOM literature, this is called the prototype. During training, the method of updating prototypes results in clustering.

\subsection{Training DESOM}
\label{sec:training}

While training the autoencoder consists of comparing the decoder output to the input, the SOM is trained by comparing the prototypes to the latent representation. 
During training, the similarity of each compressed spectrum is compared with each prototype. The most similar prototype is selected as the ``winner".  All prototypes are updated to become more similar to the compressed spectrum, but the level of update is weighted by the distance to this winning node with exponential dropoff. The updates to prototypes far from the winning node are effectively zero.
Early in training, this neighbourhood extends to many nearby nodes causing broad clustering. This neighbourhood shrinks as training continues allowing finer variation.
The rate at which this neighbourhood shrinks is a hyperparameter. 

Both component parts of the DESOM are trained together. As mentioned above, the autoencoder is trained by comparing the output of the decoder with the input spectrum using MSE. This ``loss function"  is a function of the parameters of the encoding function  $\mathbf{W_e}$, 
and the parameters of the decoding function $\mathbf{W_d}$; thus, the reconstruction loss is  $L_r(\mathbf{W_e},\mathbf{W_d})$.  The SOM loss is the MSE between the winning prototype and the input encoded spectrum.  It depends on the encoder weights ($\mathbf{W_e}$) and the weights of each prototype that determine its shape ($\left\{\mathbf{m}_k\right\}_{1\leq k \leq 225}$); thus, $L_{SOM}(\mathbf{W_e}, \mathbf{m_1},...,\mathbf{m_{225}})$. Each prototype has dimensions equal to the latent width, so our model has 225 prototypes each with 256 dimensions.

Further details of the SOM loss function are described in \citep[][and references therein]{Forest2019}.  The total loss function of the DESOM is a combination of the loss function for an autoencoder and a SOM:

\begin{equation}
\begin{multlined}
L (\mathbf{W_e}, \mathbf{W_d}, \mathbf{m_1},..., \mathbf{m_{225}}) =  \\
L_r(\mathbf{W_e},\mathbf{W_d}) +  \gamma L_{SOM}(\mathbf{W_e}, \mathbf{m_1},...,\mathbf{m_{225}})
\end{multlined}
\label{eq:loss-function}
\end{equation}

\noindent where $\gamma$ is used to tune the ratio of these two component loss functions.
For example, a $\gamma$ that is too high would mean that the ability of the autoencoder to compress a spectrum into a meaningful latent space is not prioritized enough, and meaningless data would be fed into the SOM.  A $\gamma$ that is too low would mean less priority on the SOM's ability to generate a smooth map. We have chosen $\gamma$ to be 5e$^{-4}$.
Measuring the smoothness of the map is done using ``topological error", which is the fraction of input spectra in which the ``winning" node is not adjacent to the ``runner-up" node. A well trained SOM will have a low topological error indicating that the pair of nodes most similar to the embedded input are adjacent.

To train the DESOM model, we randomly split our data into a training set of 782 389 spectra and a test set of 300,000. A test set is important to show the DESOM is robust to new spectra, and not simply memorizing the training data. This test set will be used in the following applications (\secref{sec:appfitting}).
After training, the prototypes were decoded into full spectra using the decoder part of the autoencoder. All avaialable codes can be found here
\footnote{~\href{https://github.com/finnarchinuk/GalaxyAE}{(pipeline-link)}}.

\section{Results and Potential Applications for Galaxy Research}
\label{sec:results}

\begin{figure*}
\centering
\includegraphics[width=16.cm]{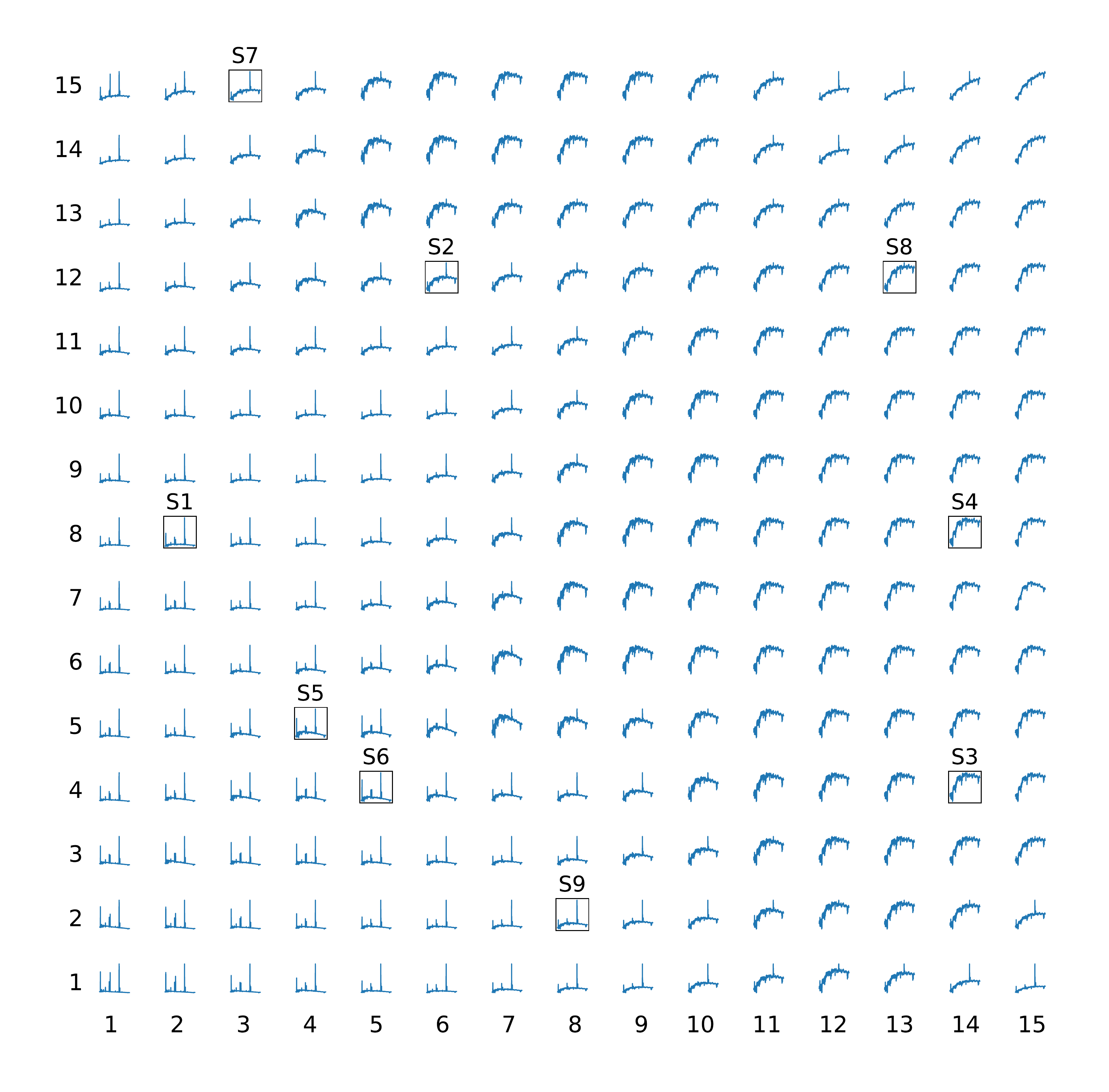}
\caption{Each node in the SOM has a corresponding prototype. The spectra shown have been generated by sending the low-dimensional prototypes from the SOM through the decoder. The sample spectra from Figure \protect\ref{fig:9spectra} have been assigned to the nodes as indicated by the label above the box.}

\label{fig:DESOM1Map-9spects}
\end{figure*}

Our trained DESOM model has organized MaNGA spectra into a 15x15 map, which is shown in Figure \ref{fig:DESOM1Map-9spects}.  The ``spectra" shown in each node is the decoded prototype representing that node. The 225 nodes can be considered an average of the MaNGA baxels that were best matched by that node.  These decoded prototypes should be treated as a diagnostic tool showing the diversity of the spectra seen by the model during the training process. As can be seen, for example, the left side of the map shows a cluster of spectra with strong emission lines and continuum shapes that are flat to blueish. In contrast, the swath on the right half of the map shows spectra that do not have significant emission lines, with continua that are red.  The spectra in the top right corner shows emission lines but red continuum shapes, and so are likely to be strongly reddened.  

To illustrate how the model organizes different spectra onto the map, we input the nine spectra shown in Figure \ref{fig:9spectra} through our model.  The boxes of Figure \ref{fig:DESOM1Map-9spects} shows the nodes to which these spectra  have been assigned. As can be seen, they are distributed on the map based on their spectral characteristics. 

In the following we demonstrate three ways in which such a map can be used in the analysis of galaxy spectra.

\subsection{An efficient alternative to full spectral fitting}
\label{sec:appfitting}

The technique of full spectral fitting consists of comparing a spectrum to a set of template spectra with known properties.  These templates are usually simple stellar populations (SSPs), i.e., populations with a single age, metallicity and alpha-enrichment.  One of the goals of full spectral fitting is to find the linear combination of SSP spectra that best fits an observed spectrum (see \citealp{Conroy2013} for a review), and to use that linear combination to infer the mass- and light-weighted properties of the stellar populations that produced the observed spectrum.  Many codes have been designed to do this, a list of public codes including pPXF can be found here\footnote{ http://www.sedfitting.org/Fitting.html}.   Some fitting codes such as Prospector \citep{Johnson2021} use Bayesian inference to compute a full probability distribution function for physical properties.

\begin{figure*}
\centering
\includegraphics[width=16.cm]{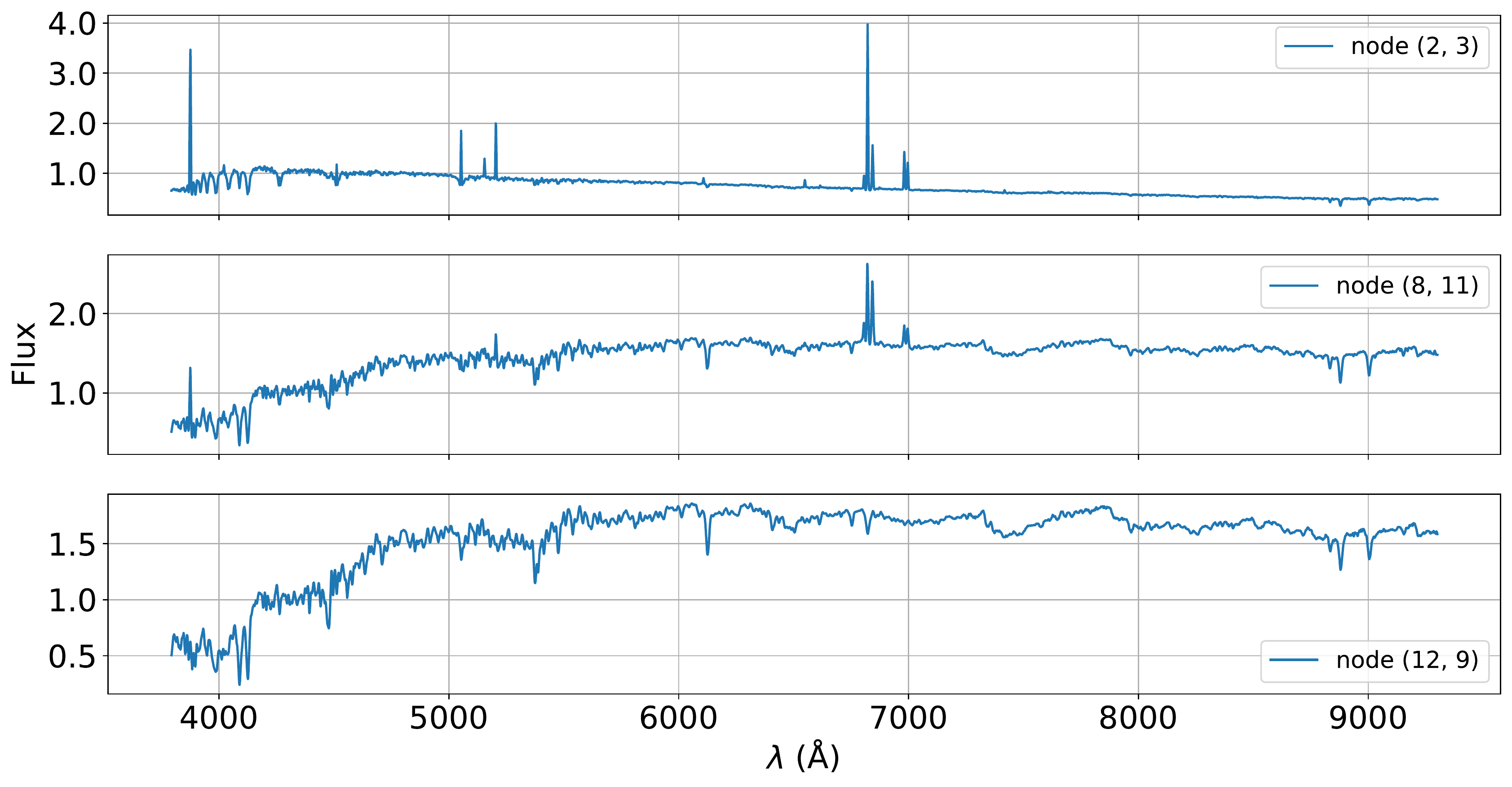}
\caption{Three template spectra for the indicated nodes. Templates were created by co-adding spectra that were assigned to the indicated node. The spectra were first normalized following \protect\secref{sec:data} before taking the mean value.}

\label{fig:sample_templates}
\end{figure*}

What we demonstrate here is an alternative to full spectral fitting by our DESOM method.  The SOM and its 225 decoded prototypes can be treated like templates, but instead of SSPs, each node is a composite of populations.  The map of templates spans the range of possible composite spectra in the local universe (as seen by MaNGA).  Figure \ref{fig:sample_templates} shows a sample of these template spectra.
 
\begin{figure*}
\centering
\includegraphics[width=18.cm,angle=0]{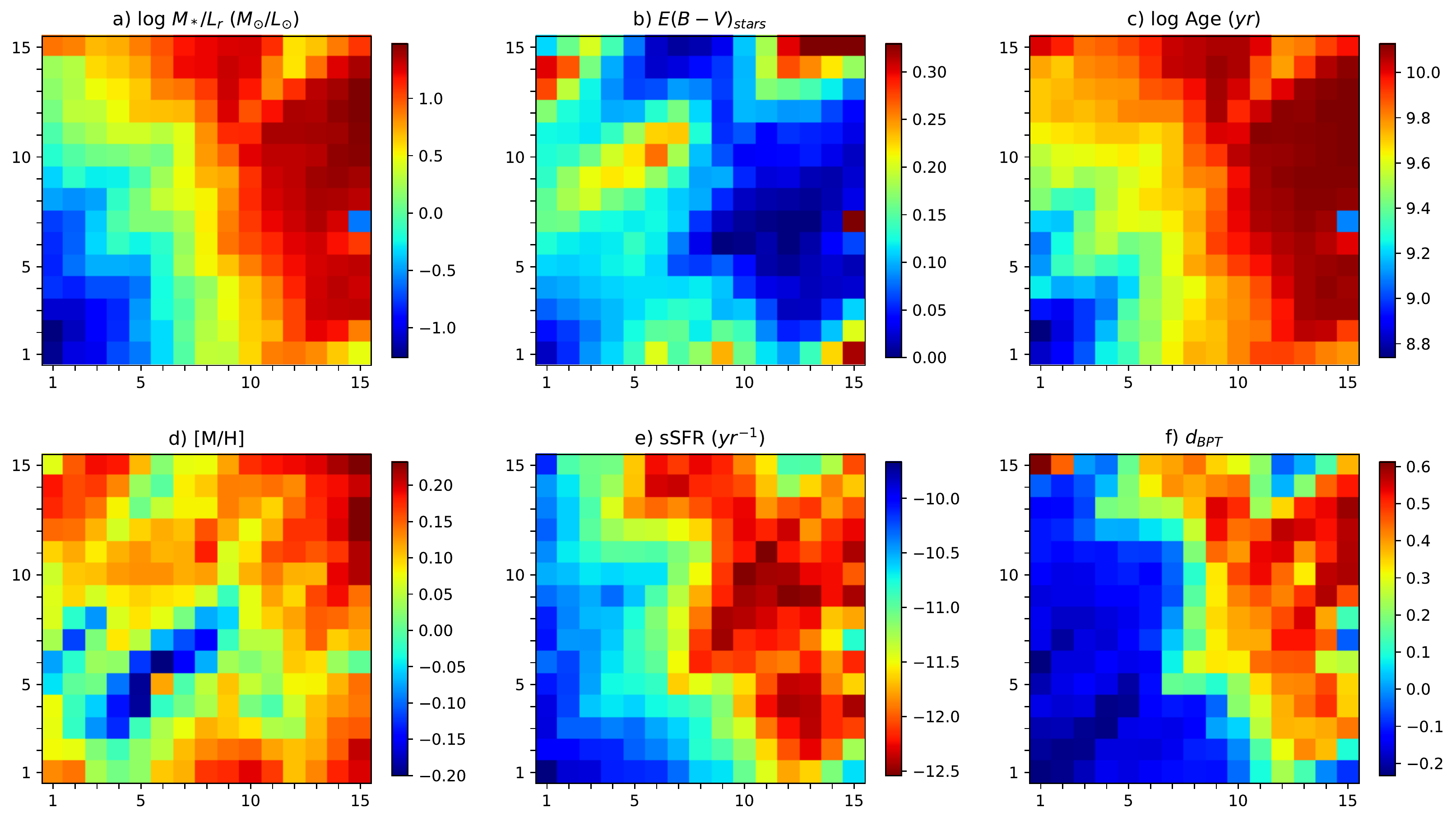}
\caption{Distribution of a selection of physical parameters from the SOM. Each node displays the median value for that parameter as determined by the baxels that were assigned to that node.  Baxels were included in the median if the reduced $\chi^2$ values of their pPXF fits were less than 2.  Note the colour scale of panel b was clipped at 3$\sigma$ of the full range.}
\label{fig:SOM1physparams}
\end{figure*}

To determine physical properties for these templates, one could take two possible approaches: use pPXF (or another full spectrum fitting code) to fit the 225 prototypes, or use pPXF to fit all the individual training spectra that were mapped onto a node.  The advantage of the first approach is the much lower number of fits necessary, but produces only point estimates for each physical parameter for each prototype. The second approach compiles point estimates from the cluster of spectra assigned to that node providing a full empirical probability distribution of a property within a single node.  We prefer the second approach, and while we have access to this full empirical probability distribution, Figure \ref{fig:SOM1physparams} shows just the median values for several physical parameters (i.e., labels) on the SOM.
These values are a combination of the training and test sets. We used the test set to ensure that the model would be robust to new spectra, and once seeing that these maps were nearly identical, we combined these two sets.

As an example, consider panel (a) of Figure \ref{fig:SOM1physparams}, which shows the map of median stellar mass-to-light ratio log $M_*/L_r$ of the baxels that were assigned into each node during training (showing only baxels whose pPXF fits had a reduced $\chi^2$ of less than 2).  It is important to note that the DESOM model never saw these physical properties at any time, and only considered the fluxes in wavelength bins.  Yet median $M_*/L_r$ varies smoothly over the SOM, and in the way we would expect: the lower left corner of the SOM, where the prototypes showed strong emission lines and blue continuum shape have low values of $M_*/L_r$, while the swath of red spectra with no emission lines have high $M_*/L_r$.  The spectra in the top right corner which we surmised were strongly reddened have intermediate values of $M_*/L_r$.  Furthermore, panel (b) of Figure \ref{fig:SOM1physparams}, which is a map of E(B-V), shows that the spectra in this region do indeed have high E(B-V).  Panel (c) shows the maps of mass-weighted stellar age which also varies over the map as expected given the prototypes.  Panel (d) shows the stellar metallicity, which does not vary as smoothly across the SOM as the other quantities.  The general trend is that the more passive spectra have higher metallicity, but many active spectra have a range of stellar metallicities.

Panels (e) and (f) of Figure \ref{fig:SOM1physparams} show maps of sSFR and $d_{\rm BPT}$.  Since these panels are derived from the emission lines, the sSFR map only includes baxels with $\Ha$ S/N $>$ 2, while the map of $d_{\rm BPT}$ includes only those baxels where the minimum S/N between the four lines of the BPT diagram ($\Hb$, [OIII]5007, $\Ha$ and [NII]6583) is greater than 2.  In general, the baxels with strong emission lines in the lower left of the maps are classified as star-forming ($d_{\rm BPT} < 0$) with high sSFR while the nodes in the passive swath have low sSFR and have AGN or LINER-like emission.

\begin{figure}
\centering
\includegraphics[width=\linewidth]{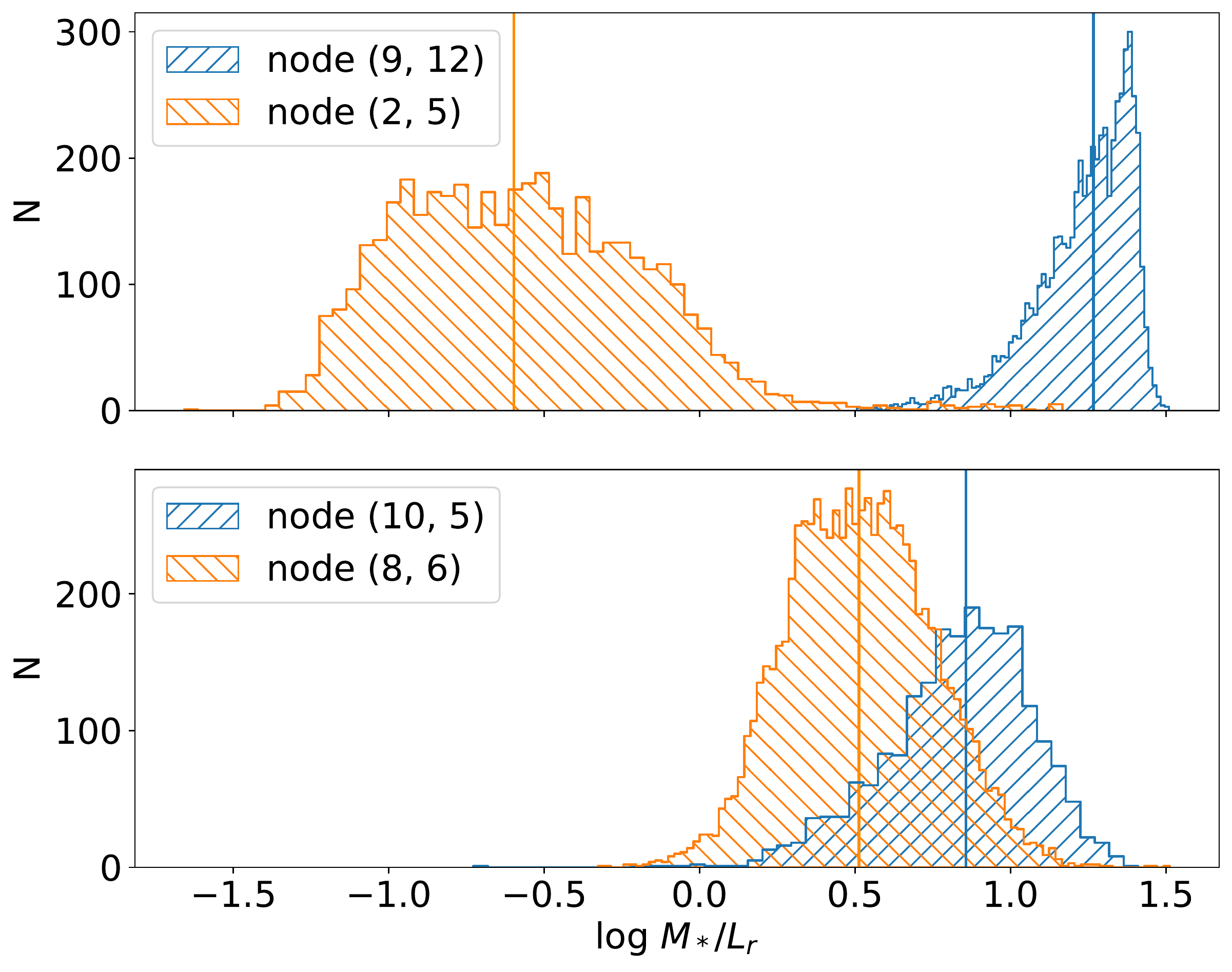}
\caption{Top: The distribution of $M_*/L_r$ values from nodes (9, 12) and (2, 5), which are widely separated on the DESOM. Older populations have higher mass to light ratios. The median values are 1.26 ± 0.51 and -0.60 ± 0.29, respectively. Bottom: The  distribution of$M_*/L_r$  in nodes are closer together on the map (10, 5) and (8, 6), with median values 0.86 ± 0.52 and 0.51 ± 0.41, respectively.}
\label{fig:mlrhist}
\end{figure}

Let us now consider the distribution of physical values within nodes.  The upper panel of Figure \ref{fig:mlrhist} shows the distributions of log $M_*/L_r$ comparing nodes (9, 12) and (2, 5) of the SOM, i.e., nodes that are widely separated on the map.  The former is from the passive swath and the latter is from the star-forming corner with the strong emission lines.  Their distributions of log $M_*/L_r$ are very different as expected, and have almost no overlap.  The lower panel compares the $M_*/L_r$ distributions in nodes (10, 5) and (8, 6) of the SOM, which are nearly adjacent.  They are both on the border between the star-forming and passive regions and their prototypes are very similar.  Figure \ref{fig:mlrhist} shows that these neighbouring nodes have large overlap in their $M_*/L_r$ distributions, but the SOM was able to distinguish these as different populations.  

The distributions of the physical properties within a node can thus be treated as an empirical probability distribution function, which can be used to infer the properties of other spectra.  When a spectrum with unknown properties is passed into the trained DESOM, the model will assign it to the most similar node, and the full empirical probability distribution of the desired properties are instantly known.  Since the DESOM is already trained, determining node position of 10000 spectra took about 150 seconds on a CPU (about 0.015 seconds per spectrum).  Compare this to one of the faster codes we are familiar with, pPXF which takes about 30-120 CPU seconds to fit a single spectrum (depending on various configuration choices), but which produces only a best-fit point estimate for each physical parameter rather than a full probability distribution. The Prospector code, which does produce a full probability distribution, takes 24-48 CPU hours to fit a single spectrum (depending on the configuration - S. Tacchella, private communication). Our method is a few thousand times faster on average than pPXF and produces a probability distribution. Furthermore our method natively runs on GPU, for additional speedup.

The empirical probability distribution of a given physical parameter is the distribution of best fit values from pPXF for the spectra in a single node, and is not the same as the actual posterior distribution of these parameters given the data. While the Bayesian exploration of the full posterior distribution is beyond the scope of this paper, our method of compiling many point estimates is akin to the Bayesian bootstrap, where we have used a sample of similar spectra as our bootstrap sample. While this method of compiling point estimates is not exactly the same as the Bayesian bootstrap, such methods have been shown to simulate the posterior reasonably well \citep[e.g.,][]{Rubin81} so we are confident that the empirical probability distributions are close to the actual posteriors, given the quality of the data.  

It should be reiterated that the SOM does not operate on raw spectra, but rather on their latent representation from the autoencoder. Therefore our empirical probability distribution is P(physical parameter $|$ latent representations) rather than P(physical parameter $|$ data). While the autoencoder constitutes a significant non-linear transformation of our data, similar spectra should have similar latent representations because we are using an undercomplete autoencoder and the model maintains a low reconstruction error. Therefore we do not expect the probability P(physical parameter $|$ data) to be very different from P(physical parameter $|$ latent representations). 

One clear disadvantage of our method as an alternative to full spectral fitting is the need for a training set, and the prior fitting of this training set using traditional codes like pPXF. However, once we have a training set, it takes only 4 hours to train the DESOM. From that point onwards, new spectra can very quickly be assigned a label.
While this setup time may seem significant, given the vastness of past and future spectroscopic surveys involving $10^{6-8}$ spectra, the need for ML alternatives to full spectral fitting, such as our DESOM method, becomes clear.

\subsection{Mapping the IFU spectra from a single galaxy}
\label{sec:appifu}

The use of the DESOM map as templates as described above could in principle be applied to any survey of spectra, including stellar libraries, or spectroscopic surveys of galaxies such as the Sloan Digital Sky Survey, where a single spectrum is obtained per object.  Here we demonstrate how the DESOM map can be used to study IFU data, where multiple spectra are measured across many positions in a galaxy.  

\begin{figure*}
\centering
\includegraphics[width=16.cm]{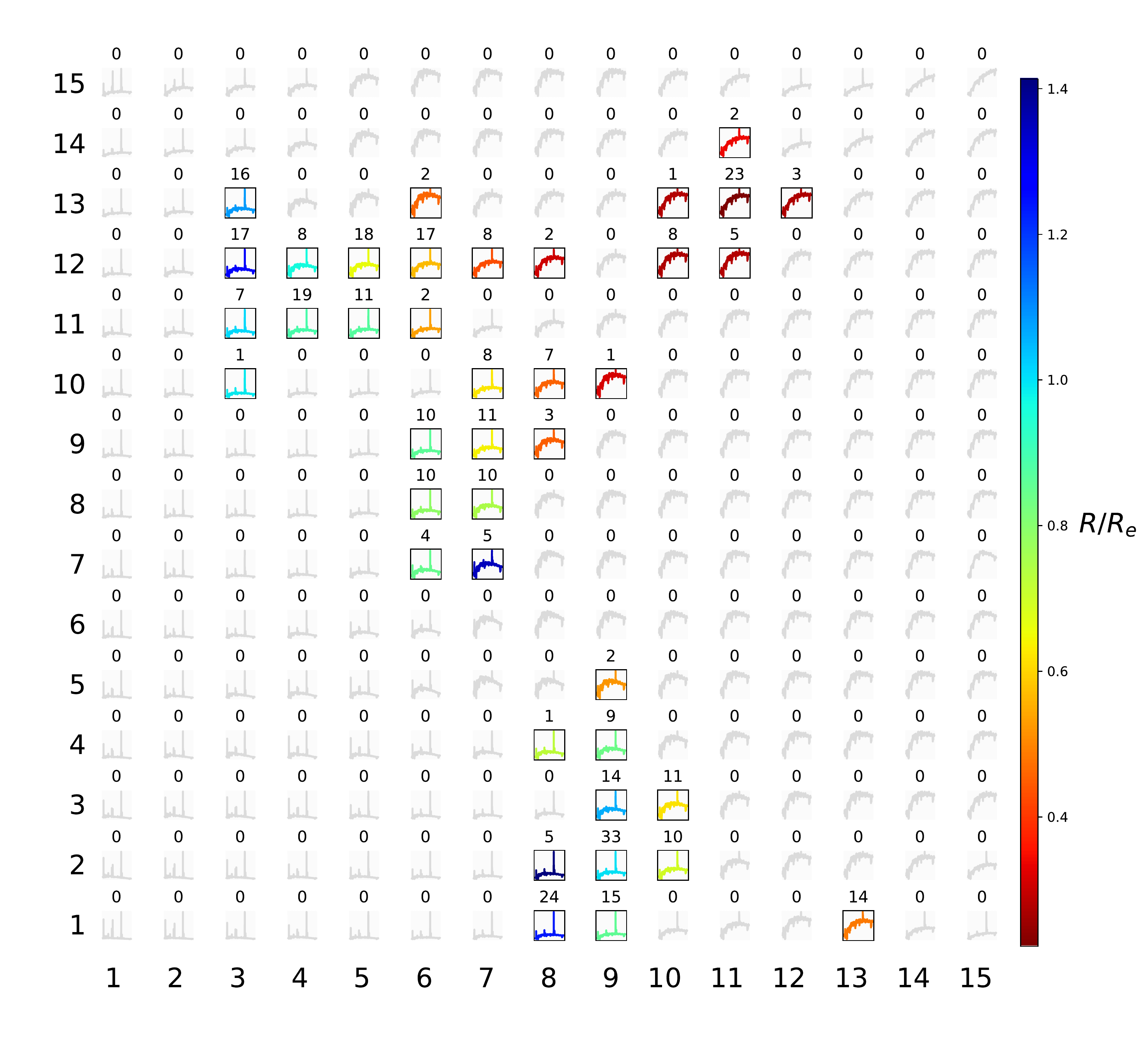}
\caption{By assigning all baxel spectra from the MaNGA galaxy 1-114956 to the SOM, we can summarize this galaxy. The number above each decoded prototype indicates how many spectra have been assigned to the node. These decoded prototypes have been coloured by mean half-light radius.}
\label{fig:fig-som1-galaxy}
\end{figure*}

The distribution of baxels taken from a single MaNGA galaxy onto the DESOM map has significant potential for galaxy research.  As an example, we input the baxels from a selected galaxy (MaNGA ID: 1-114956) to the model and show their distribution onto the SOM in Figure \ref{fig:fig-som1-galaxy}.  Numbers above each prototype indicate the number of baxels that are best represented by that node, and the prototypes are colour-coded by their galactocentric radius in units of the half-light radius.  This galaxy appears to have several significant groupings of baxels.  The ones near node (11, 13) are in the centre of the galaxy with older spectra and LINER/AGN-like emission lines.  The ones near (6,12) are at intermediate-to-large radii are a group with both older and younger spectra.  The group of baxels near (9, 2) are at mostly large radii and more strongly star-forming.  Since these groupings of baxels are widely separated, this suggests an episodic star-formation history for this galaxy, or recent merger or accretion events. 

This use case, namely studying the distribution of the baxels of a galaxy, builds on the above model. By looking at the entire set of spectra from a galaxy at once, rather than one at a time, we can start to uncover information about the target galaxy as a whole. The DESOM model still does not take spatial information into consideration; we are applying labels only after node assignments have been made.

\subsection{DESOM for large galaxy IFU surveys}
\label{sec:appd2}

Clearly, different galaxies will populate the DESOM map differently, forming a unique ``fingerprint" of baxels on the map.  Spiral galaxies will have many baxels occupying the lower left corner of the map, while elliptical galaxies might primarily occupy the passive swath.  Given the 4609 galaxies currently available in MaNGA, how can these fingerprints be summarized in an efficient way so as to extract scientific value?  Here we demonstrate a method to use yet another SOM to organize {\it galaxies} by their IFU fingerprints on the DESOM map.
This is similar to, but distinct from, \cite{Sarmiento} who have applied contrastive learning to the MaNGA dataset using 2-dimensional maps of derived physical parameters to cluster galaxies. The maps they chose extract important features from associated spectra and maintains the 2-dimensional spatial information of the galaxy. Our method generates a single map that removes the spatial information, but allows direct reconstruction of the original spectra.

Above, we examined the fingerprint of MaNGA galaxy 1-114956 in Figure \ref{fig:fig-som1-galaxy} colour-coded by median galactocentric radius.  In the same figure, the numbers above each decoded prototype shows the numbers of baxels from this galaxy that are represented by that node. For example, node (9, 2) has 33 `similar' spectra, whereas node (12, 7) has zero members. 
Note that due to differing IFU sizes used in MaNGA, and due to differing S/N between cubes, our Voronoi binning (described in \secref{sec:data}) resulted in different numbers of baxels for each galaxy.  
Regardless of the number of baxels in a galaxy, the map is essentially a 2D histogram of its baxels.

\begin{figure*}
\centering
\includegraphics[width=16.cm]{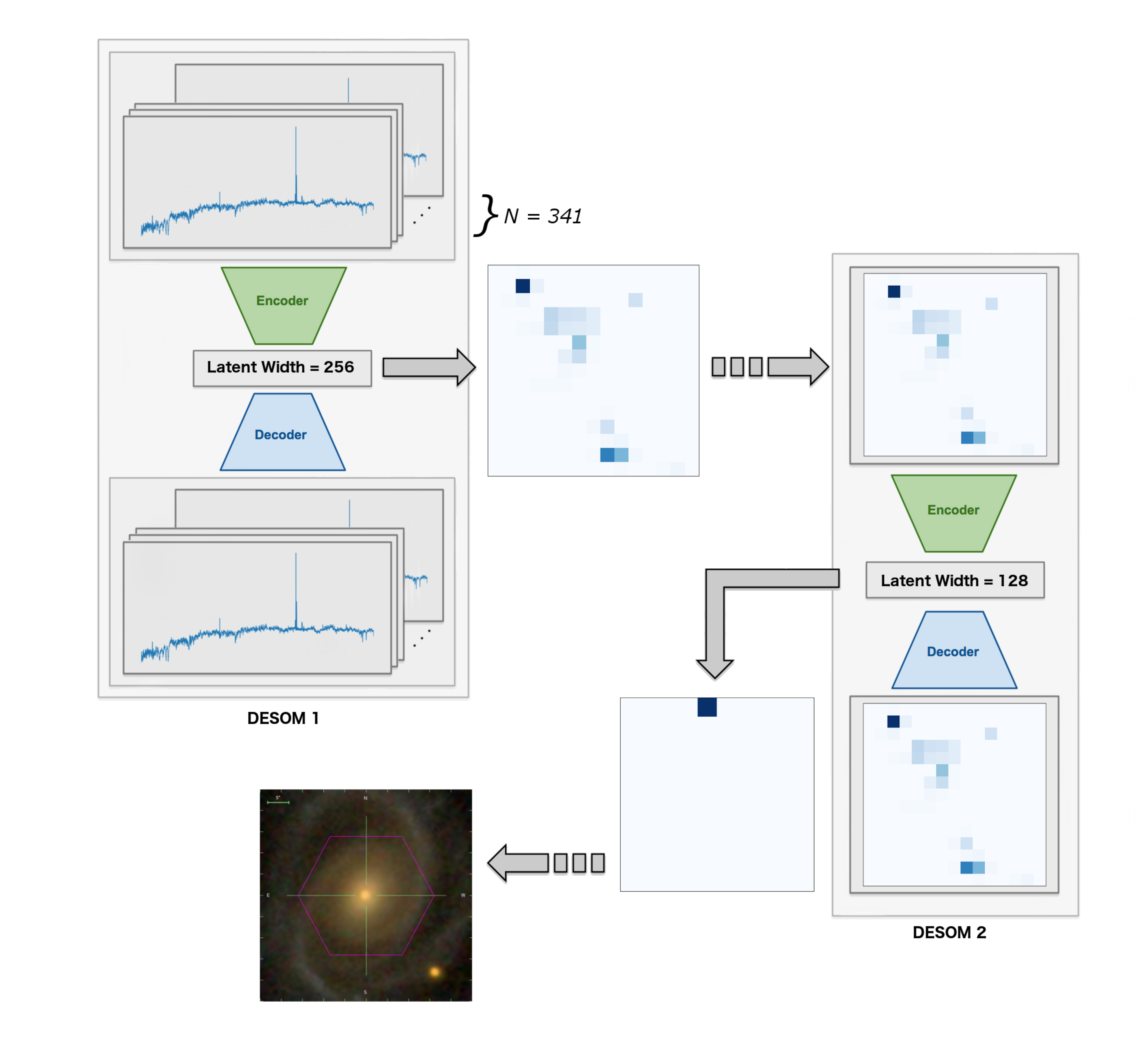}
\caption{As an example, 341 spectra of a galaxy are reduced to a width of 256 by an autoencoder. The 341 spectra are distributed on the map as a unique pattern being the fingerprint for that galaxy.
The unique patterns of all galaxies in the survey can be fed into another DESOM. The output of this DESOM is a distribution of galaxies clustered by similarity.
This example galaxy is assigned to the top middle edge of DESOM-2, with an image shown.
}
\label{fig:pipeline}
\end{figure*}

\cite{Teimoorinia21b} explored using sequential DESOMs to cluster massive complex images. We use a similar technique of sequential DESOMs to cluster galaxies by their spectra (as summarized by DESOM-1). The inputs for DESOM-2 will be the 2D maps of galaxies from DESOM-1.
Figure \ref{fig:pipeline} illustrates our pipeline for organizing galaxies, using the MaNGA galaxy 1-122088 and is 341 baxels as an example.

Given the dimensionality of these fingerprints, galaxies could be clustered directly using a simple SOM, bypassing the autoencoder. We chose to use a full DESOM for the following reasons:
a SOM will work better if the inputs are information dense (the fingerprints are sparse with the majority of nodes unused for any given galaxy),
and we are limited to $<$5000 galaxies, so compressing the fingerprints is a form of regularization.
To avoid the \textit{curse of dimensionality} problem \citep{bishop:2006}, it is important to increase the number of samples, or decrease the dimensionality of the data. Since we have a fixed number of galaxies available, we use the autoencoder to further reduce the complexity of our fingerprints.

With only 4609 galaxies available,  we trained on 80$\%$, and tested on 20$\%$. For each node in the training set, the node was clipped to the 99.9th percentile, then scaled from 0 to 1. This normalization weights the nodes by importance given the observed variation. Then each fingerprint was scaled to between 0 and 1 so the ML model has consistently scaled samples which helps in training. We compare the resulting maps from the training and test sets using their distributions on SOM2, and reconstruction errors. From this analysis we conclude there is no significant difference between them. Since they are comparable our following results use values from the entire set.
We trained this model until the MSE of the validation set plateaued during training.  

\begin{figure}
\centering
\includegraphics[width=\linewidth,angle=0]{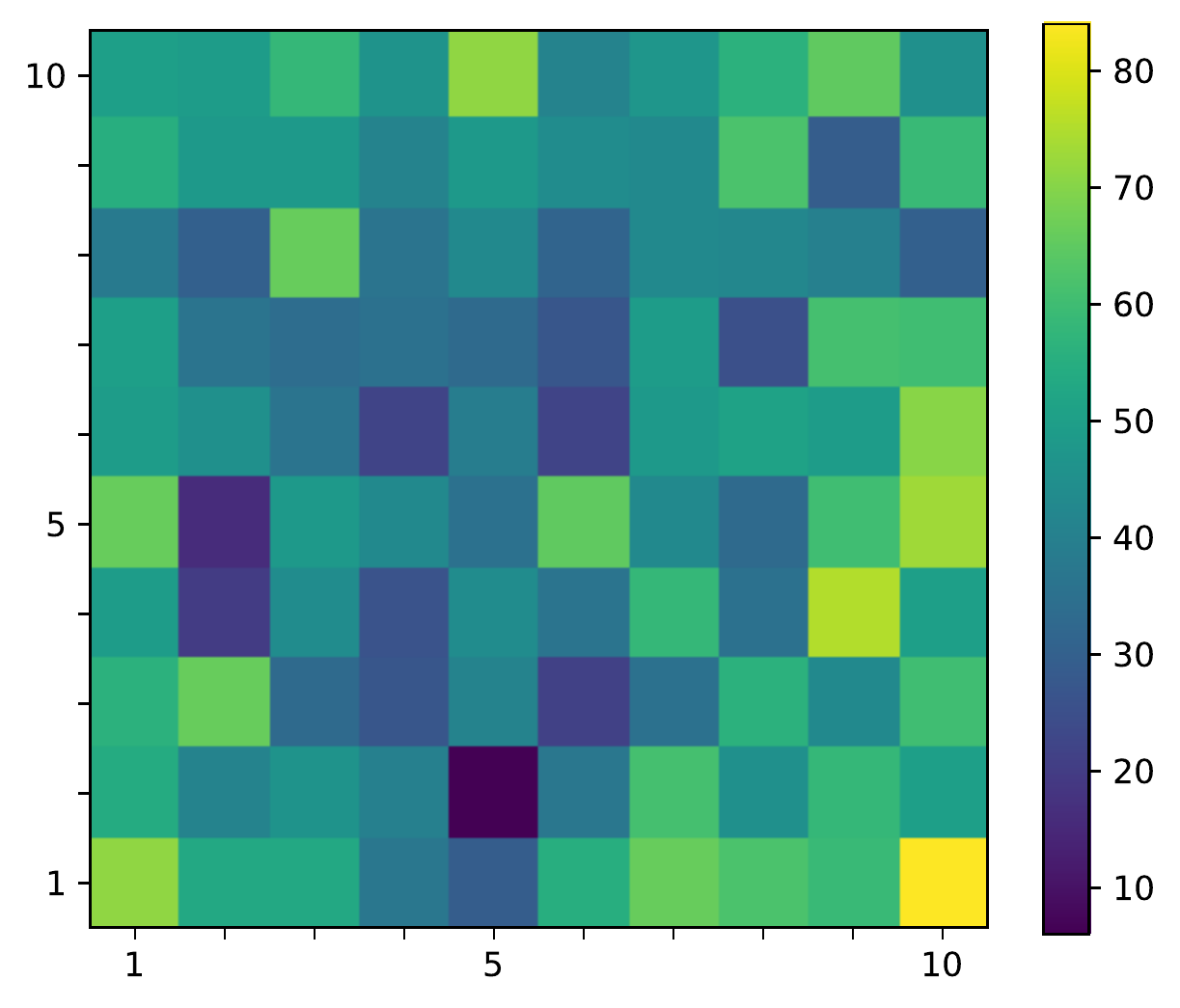}
\caption{The number of galaxies assigned to each node of DESOM-2 show the map is well utilized.}
\label{fig:d2usage}
\end{figure}

In Figure \ref{fig:d2usage} we show the resulting 10x10 map colour-coded by the number of galaxies in each node. Node (5, 2) is the least populated node because the adjacent nodes (4, 2) and (6, 2) are irregular and elliptical galaxies. Since a SOM updates the winning node and the immediate neighbours, this node is pulled into a physically meaningless intermediate state. This marks the beginning of a separation boundary where a SOM forgoes using some nodes to split the map into distinct regions.

\begin{figure*}
\centering
\includegraphics[height=16.cm,angle=0]{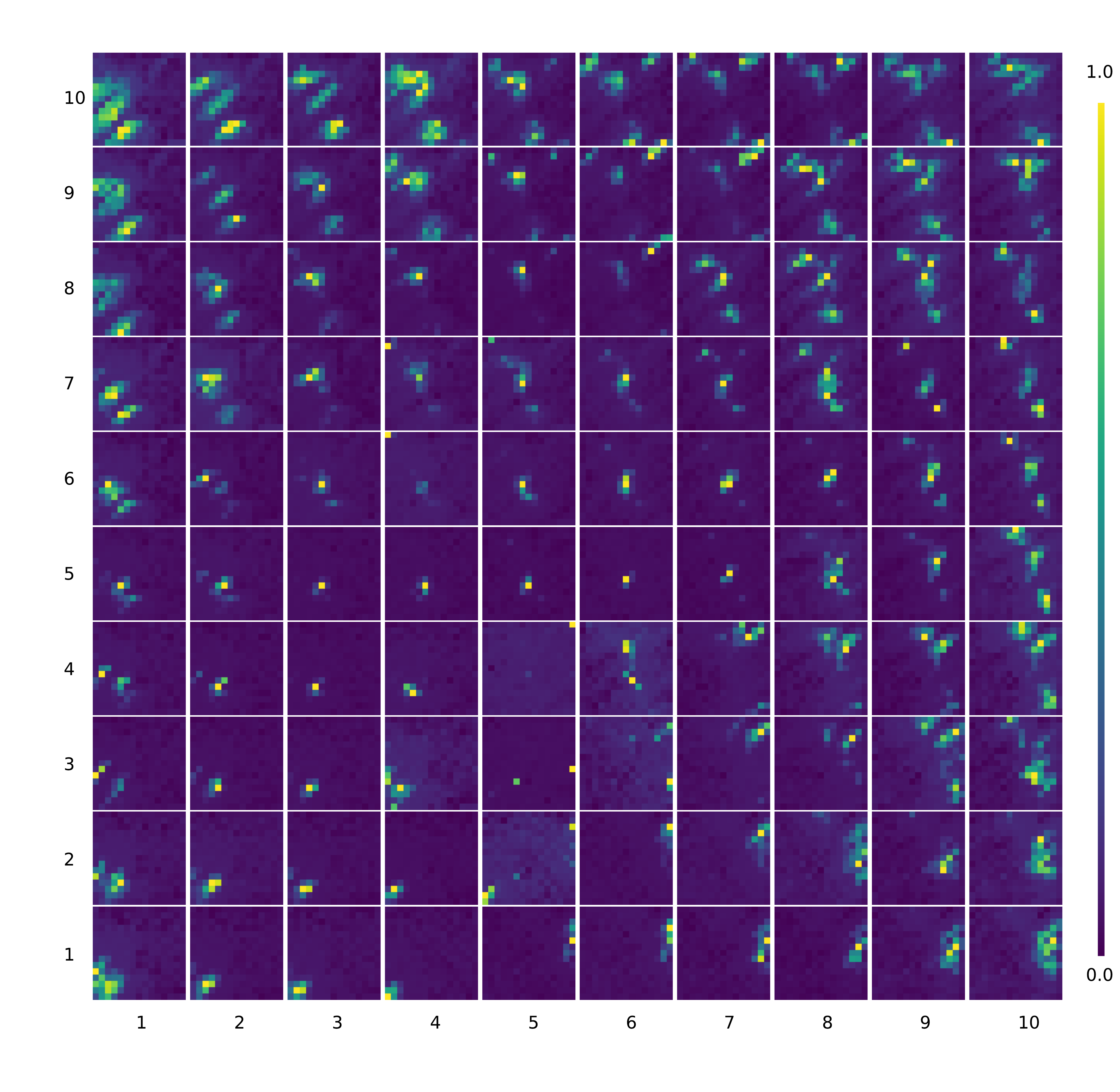}
\caption{Decoded prototype  galaxies of DESOM-2. DESOM-2 assigns each galaxy to one of these nodes based on how similar the fingerprint is to the prototype. Elliptical galaxies have spectral characteristics that cause the fingerprint to be clustered into the bottom right corner (see Figure \protect\ref{fig:DESOM2_images}).}
\label{fig:DESOM2_protos}
\end{figure*}

The resulting map of decoded fingerprint prototypes can be found in Figure \ref{fig:DESOM2_protos}, which represents the full diversity of galaxy fingerprints within MaNGA. Each prototype is the average fingerprint of galaxies in that node, and are colour-coded by the relative number of baxels that fall in that region of the DESOM-1 map, i.e., the ``activation" of the nodes in DESOM-1. For example, the prototype fingerprint at node (10, 1) represents galaxies whose baxels occupy the middle right side of DESOM-1.  Referring back to Figure \ref{fig:DESOM1Map-9spects}, this is the region of DESOM-1 where baxels have are passive.  Since DESOM-1 grouped the spectra by similarlity, the smoothness and contiguity of the fingerprint in this node indicates a smoothly varying stellar population over the spatial extent of the galaxies in this node.  These galaxies likely experienced a smoothly varying star-formation history and/or relatively quiet merger history.  Many of the other prototypes such as (7, 10) show multiple activated patches in widely separated regions.  Again, since the baxels are grouped onto the fingerprints by similarity, the multiple activated regions indicate multiple stellar populations, suggesting a more episodic star-formation history or major merger.

\begin{figure*}
\centering
\includegraphics[height=16.cm,angle=0]{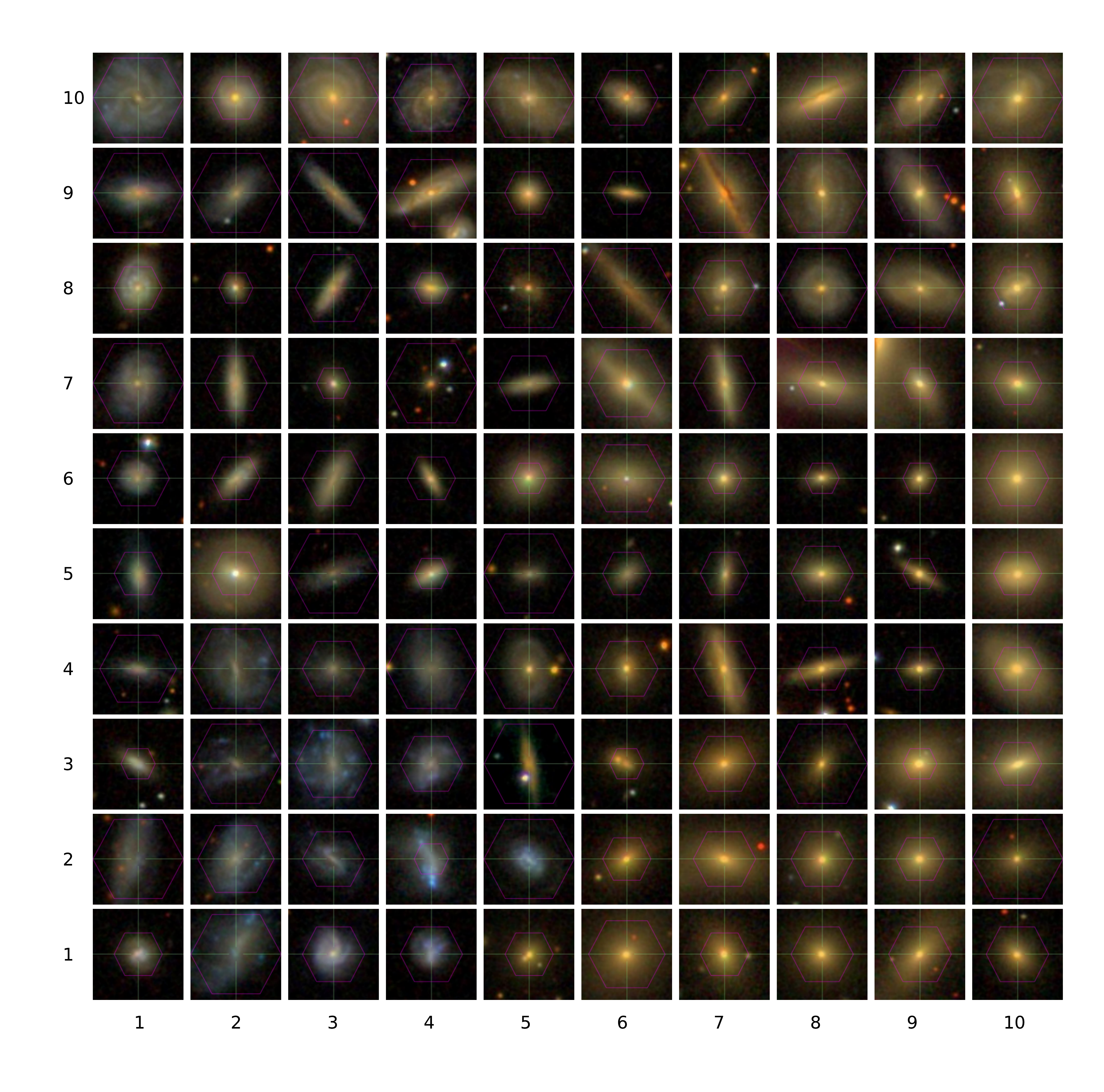}
\caption{Each of the 4609 galaxies is assigned a node on DESOM-2. Here we show one randomly selected galaxy from each node to show the general characteristic of that node. The number of galaxies in each node can be found in Figure \protect\ref{fig:d2usage}.}
\label{fig:DESOM2_images}
\end{figure*}

\begin{figure*}
\centering
\includegraphics[width=16.cm,angle=0]{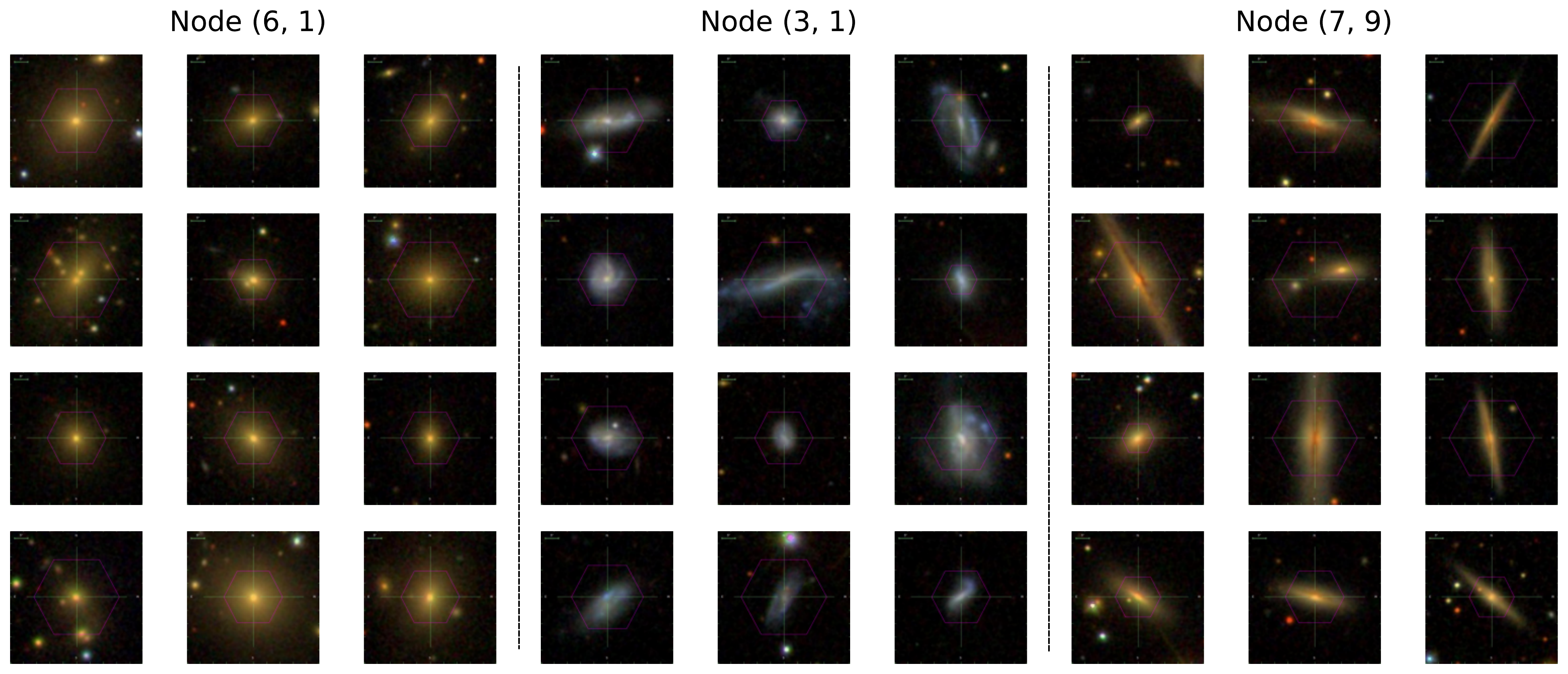}
\caption{Three nodes have been selected for additional scrutiny. The first three columns are from node (6, 1), the next three columns are from node (3, 1) and finally from (7, 9).}
\label{fig:zoomnodes}
\end{figure*}

What kinds of galaxies do the fingerprints in the DESOM-2 map represent?  We randomly select one galaxy from each node in DESOM-2 and download their images using the MaNGA collaboration's Marvin API \citep{Cherinka2019}.  These images are shown in Figure \ref{fig:DESOM2_images}.  From this figure we see clear morphological divisions between certain regions of the map, with elliptical galaxies being clustered in the lower right.
To show that similar galaxies are assigned to the same node in the DESOM-2 map, we randomly selected 12 galaxies from nodes (6, 1), (3, 1), and (7, 9), which each have at least 40 galaxies assigned to them, and show their images in Figure \ref{fig:zoomnodes}. This figure shows that the galaxies within each node are morphologically similar.

\begin{figure*}
\centering
\includegraphics[width=18.cm,angle=0]{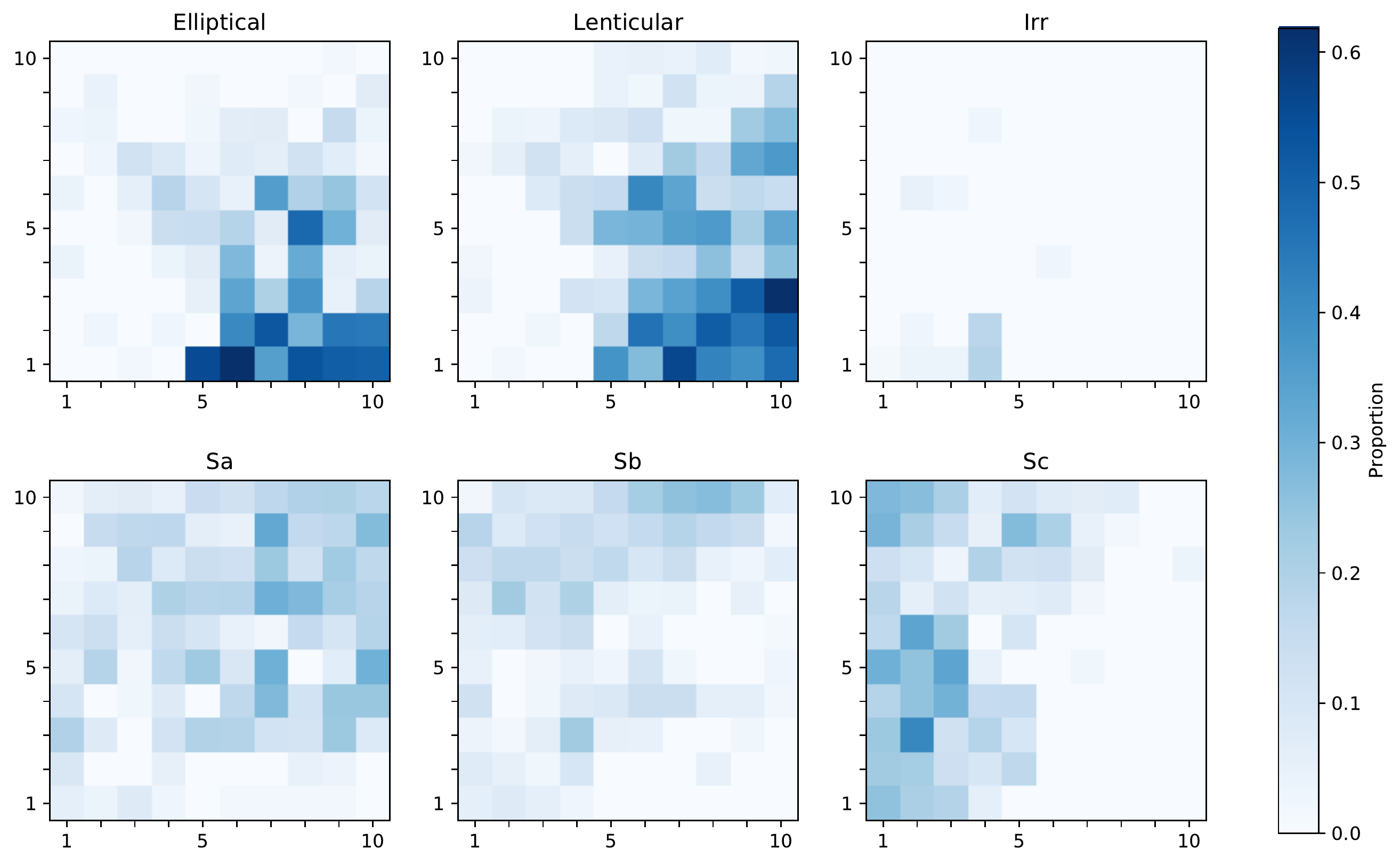}
\caption{We select a morphology classification and show the proportion of galaxies in each node that are assigned that classification.
Morphology information is from \protect\cite{HernandezToledo2010}}
\label{fig:d2morphologies}
\end{figure*}

We also divided the galaxies by visual morphological classification (from \citealp{HernandezToledo2010}) and show their maps in Figure \ref{fig:d2morphologies}.  This figure shows that certain morphological types occupy different regions of the DESOM-2 map.  

\begin{figure*}
\centering
\includegraphics[width=16.cm]{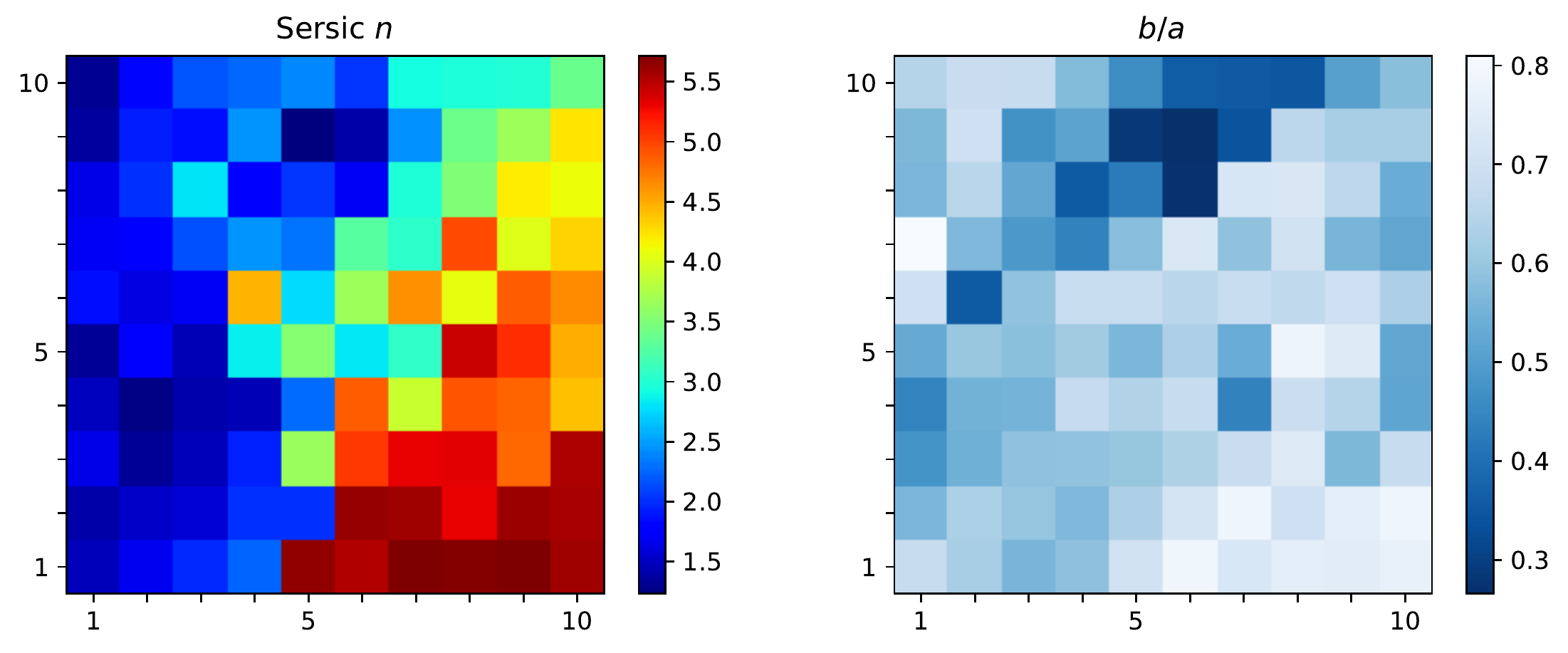}
\caption{Sersic and b/a values for each node of DESOM-2 based on the galaxies that were assigned there.}
\label{fig:DESOM2_params}
\end{figure*}

Lastly, in Figure \ref{fig:DESOM2_params} we show maps of the median values of two other morphological measurements from the r-band Sersic fits of the MaNGA Data Reduction Pipeline (v. 2.4.3; \citealp{Law2016}): Sersic $n$ and the r-band axial ratio $b/a$.  Regions of high and low Sersic $n$, as well as axial ratio are clearly separated in the DESOM-2 map.  Even edge-on galaxies are grouped together on the map, being dominant in a swath from around (7,10) to (2,6).

It is important to emphasize that information about the spatial distribution of baxels within a galaxy was not provided to either DESOM model.  Only the numbers of baxels of a certain type (i.e., the galaxy fingerprints on DESOM-1) were provided to DESOM-2.  Thus the results of DESOM-2 combined with information that was withheld from the models (in this case, morphology) can be used to make scientific inferences.  For example, many of the spiral galaxies, as well as some early-type galaxies, appear to have multiple stellar populations.

\section{Summary and Conclusions}
\label{sec:summary}

We have presented a deep unsupervised machine learning method (DESOM) for organizing and labelling astronomical spectra and demonstrated its power for scientific discovery. We applied our methods to spectra from the MaNGA IFU galaxy survey, showing that the entire diversity of MaNGA spectra can be organized onto a 15x15 map (DESOM-1 - Figure \ref{fig:DESOM1Map-9spects}), where neighbouring points on the map represent similar spectra.  We demonstrated how this map can be used as an alternative to conventional methods of full spectral fitting for deriving physical quantities from the spectra (such as the stellar mass-to-light ratio, stellar age and metallicity, as well as emission line quantities such as the sSFR and BPT diagnostics).  We showed how our method can be used to empirically derive full probability distributions for these quantities much more efficiently than traditional methods.  

Since we have applied our method to the MaNGA survey, we can take advantage of spatial information to examine the distribution of spectra onto the DESOM map for a single galaxy, i.e., its DESOM ``fingerprint".  Since spectra are grouped by similarity, the distribution of spectra onto the map yields information about the stellar (and gas) populations within the galaxy.  Contiguous activation regions on the map indicate smoother star-formation histories and quieter merger histories, while a galaxy with spectra that are widely separated on the map experienced a more episodic star-formation history or major merger events.  

Since the latest public data release of MaNGA (DR15) includes 4609 galaxies, we used a second DESOM to map the diversity of galaxy fingerprints (DESOM-2 - Figure \ref{fig:DESOM2_protos}).  Examining the images of the galaxies, as well as several independent measures of galaxy morphology, we confirm that galaxies with similar DESOM-1 fingerprints also have similar morphologies.  Even edge-on galaxies are grouped together in DESOM-2.  Since morphological and spatial information was not provided to either DESOM, linking galaxies with a certain morphology with the star-formation histories encoded in their fingerprints is one example of how the DESOM maps can be used to make scientific inferences.

Our analysis has not included positional information of baxels within galaxies.  Only the numbers of baxels in different nodes from DESOM-1 were provided as input for DESOM-2.  So, while we have demonstrated, for example, how one might detect multiple distinct populations within a galaxy, where they are in the galaxy is not discerned by DESOM-2.  One promising line of inquiry for future studies would be to devise a method for weighting the fingerprint regions by galactocentric distance before feeding them to a second DESOM.  It is clear that DESOMs have significant potential for aiding galaxy evolution research.

\bibliography{main,references}{}

\begin{thebibliography}{}
\expandafter\ifx\csname natexlab\endcsname\relax\def\natexlab#1{#1}\fi
\providecommand{\url}[1]{\href{#1}{#1}}
\providecommand{\dodoi}[1]{doi:~\href{http://doi.org/#1}{\nolinkurl{#1}}}
\providecommand{\doeprint}[1]{\href{http://ascl.net/#1}{\nolinkurl{http://ascl.net/#1}}}
\providecommand{\doarXiv}[1]{\href{https://arxiv.org/abs/#1}{\nolinkurl{https://arxiv.org/abs/#1}}}

\bibitem[{Baldwin {et~al.}(1981)Baldwin, Phillips, \& Terlevich}]{Baldwin1981}
Baldwin, J.~A., Phillips, M.~M., \& Terlevich, R. 1981, Publications of the
  Astronomical Society of the Pacific, 93, 5, \dodoi{10.1086/130766}

\bibitem[{{Bickley} {et~al.}(2021){Bickley}, {Bottrell}, {Hani}, {Ellison},
  {Teimoorinia}, {Yi}, {Wilkinson}, {Gwyn}, \& {Hudson}}]{Bickley21}
{Bickley}, R.~W., {Bottrell}, C., {Hani}, M.~H., {et~al.} 2021, \mnras, 504,
  372, \dodoi{10.1093/mnras/stab806}

\bibitem[{Bishop(2006)}]{bishop:2006}
Bishop, C.~M. 2006, Pattern Recognition and Machine Learning (Springer)

\bibitem[{{Bluck} {et~al.}(2020){Bluck}, {Maiolino}, {S{\'a}nchez}, {Ellison},
  {Thorp}, {Piotrowska}, {Teimoorinia}, \& {Bundy}}]{Bluck20}
{Bluck}, A. F.~L., {Maiolino}, R., {S{\'a}nchez}, S.~F., {et~al.} 2020, \mnras,
  492, 96, \dodoi{10.1093/mnras/stz3264}

\bibitem[{{Bottrell} {et~al.}(2019){Bottrell}, {Hani}, {Teimoorinia},
  {Ellison}, {Moreno}, {Torrey}, {Hayward}, {Thorp}, {Simard}, \&
  {Hernquist}}]{Bottrell19}
{Bottrell}, C., {Hani}, M.~H., {Teimoorinia}, H., {et~al.} 2019, \mnras, 490,
  5390, \dodoi{10.1093/mnras/stz2934}

\bibitem[{Bundy {et~al.}(2015)Bundy, Bershady, Law, Yan, Drory, MacDonald,
  Wake, Cherinka, S{\'{a}}nchez-Gallego, Weijmans, Thomas, Tremonti, Masters,
  Coccato, Diamond-Stanic, Arag{\'{o}}n-Salamanca, Avila-Reese, Badenes,
  Falc{\'{o}}n-Barroso, Belfiore, Bizyaev, Blanc, Bland-Hawthorn, Blanton,
  Brownstein, Byler, Cappellari, Conroy, Dutton, Emsellem, Etherington,
  Frinchaboy, Fu, Gunn, Harding, Johnston, Kauffmann, Kinemuchi, Klaene,
  Knapen, Leauthaud, Li, Lin, Maiolino, Malanushenko, Malanushenko, Mao,
  Maraston, McDermid, Merrifield, Nichol, Oravetz, Pan, Parejko, Sanchez,
  Schlegel, Simmons, Steele, Steinmetz, Thanjavur, Thompson, Tinker, Bosch,
  Westfall, Wilkinson, Wright, Xiao, \& Zhang}]{Bundy2015}
Bundy, K., Bershady, M.~A., Law, D.~R., {et~al.} 2015, The Astrophysical
  Journal, 798, 7, \dodoi{10.1088/0004-637X/798/1/7}

\bibitem[{Calzetti {et~al.}(2000)Calzetti, Armus, Bohlin, Kinney, Koornneef, \&
  Storchi‐Bergmann}]{Calzetti2000}
Calzetti, D., Armus, L., Bohlin, R.~C., {et~al.} 2000, The Astrophysical
  Journal, 533, 682, \dodoi{10.1086/308692}

\bibitem[{Cappellari(2017)}]{Cappellari2017}
Cappellari, M. 2017, Monthly Notices of the Royal Astronomical Society, 466,
  798, \dodoi{10.1093/mnras/stw3020}

\bibitem[{Cappellari \& Copin(2003)}]{Cappellari2003a}
Cappellari, M., \& Copin, Y. 2003, Monthly Notices of the Royal Astronomical
  Society, 342, 345, \dodoi{10.1046/j.1365-8711.2003.06541.x}

\bibitem[{Cardelli {et~al.}(1989)Cardelli, Clayton, \& Mathis}]{Cardelli1989}
Cardelli, J.~A., Clayton, G.~C., \& Mathis, J.~S. 1989, The Astrophysical
  Journal, 345, 245, \dodoi{10.1086/167900}

\bibitem[{Cherinka {et~al.}(2019)Cherinka, Andrews, S{\'{a}}nchez-Gallego,
  Brownstein, Argudo-Fern{\'{a}}ndez, Blanton, Bundy, Jones, Masters, Law,
  Rowlands, Weijmans, Westfall, \& Yan}]{Cherinka2019}
Cherinka, B., Andrews, B.~H., S{\'{a}}nchez-Gallego, J., {et~al.} 2019, The
  Astronomical Journal, 158, 74, \dodoi{10.3847/1538-3881/ab2634}

\bibitem[{Cid~Fernandes {et~al.}(2005)Cid~Fernandes, Mateus, Sodr{\'{e}},
  Stasi{\'{n}}ska, \& Gomes}]{CidFernandes2005}
Cid~Fernandes, R., Mateus, A., Sodr{\'{e}}, L., Stasi{\'{n}}ska, G., \& Gomes,
  J.~M. 2005, Monthly Notices of the Royal Astronomical Society, 358, 363,
  \dodoi{10.1111/j.1365-2966.2005.08752.x}

\bibitem[{Conroy(2013)}]{Conroy2013}
Conroy, C. 2013, Annual Review of Astronomy and Astrophysics, 51, 393,
  \dodoi{10.1146/annurev-astro-082812-141017}

\bibitem[{Croom {et~al.}(2012)Croom, Lawrence, Bland-Hawthorn, Bryant, Fogarty,
  Richards, Goodwin, Farrell, Miziarski, Heald, Jones, Lee, Colless, Brough,
  Hopkins, Bauer, Birchall, Ellis, Horton, Leon-Saval, Lewis,
  L{\'{o}}pez-S{\'{a}}nchez, Min, Trinh, \& Trowland}]{Croom2012}
Croom, S.~M., Lawrence, J.~S., Bland-Hawthorn, J., {et~al.} 2012, Monthly
  Notices of the Royal Astronomical Society, 421, 872,
  \dodoi{10.1111/j.1365-2966.2011.20365.x}

\bibitem[{{Drory} {et~al.}(2015){Drory}, {MacDonald}, {Bershady}, {Bundy},
  {Gunn}, {Law}, {Smith}, {Stoll}, {Tremonti}, {Wake}, {Yan}, {Weijmans},
  {Byler}, {Cherinka}, {Cope}, {Eigenbrot}, {Harding}, {Holder}, {Huehnerhoff},
  {Jaehnig}, {Jansen}, {Klaene}, {Paat}, {Percival}, \& {Sayres}}]{Drory15}
{Drory}, N., {MacDonald}, N., {Bershady}, M.~A., {et~al.} 2015, \aj, 149, 77,
  \dodoi{10.1088/0004-6256/149/2/77}

\bibitem[{Fitzpatrick(1999)}]{Fitzpatrick1999}
Fitzpatrick, E.~L. 1999, Publications of the Astronomical Society of the
  Pacific, 111, 63, \dodoi{10.1086/316293}

\bibitem[{Forest {et~al.}(2019)Forest, Lebbah, Azzag, \& Lacaille}]{Forest2019}
Forest, F., Lebbah, M., Azzag, H., \& Lacaille, J. 2019, in European Symposium
  on Artificial Neural Networks, Computational Intelligence and Machine
  Learning (ESANN 2019), 1--6

\bibitem[{{Gallazzi} {et~al.}(2006){Gallazzi}, {Charlot}, {Brinchmann}, \&
  {White}}]{Gallazzi06}
{Gallazzi}, A., {Charlot}, S., {Brinchmann}, J., \& {White}, S. D.~M. 2006,
  \mnras, 370, 1106, \dodoi{10.1111/j.1365-2966.2006.10548.x}

\bibitem[{{Gunn} {et~al.}(2006){Gunn}, {Siegmund}, {Mannery}, {Owen}, {Hull},
  {Leger}, {Carey}, {Knapp}, {York}, {Boroski}, {Kent}, {Lupton}, {Rockosi},
  {Evans}, {Waddell}, {Anderson}, {Annis}, {Barentine}, {Bartoszek}, {Bastian},
  {Bracker}, {Brewington}, {Briegel}, {Brinkmann}, {Brown}, {Carr},
  {Czarapata}, {Drennan}, {Dombeck}, {Federwitz}, {Gillespie}, {Gonzales},
  {Hansen}, {Harvanek}, {Hayes}, {Jordan}, {Kinney}, {Klaene}, {Kleinman},
  {Kron}, {Kresinski}, {Lee}, {Limmongkol}, {Lindenmeyer}, {Long}, {Loomis},
  {McGehee}, {Mantsch}, {Neilsen}, {Neswold}, {Newman}, {Nitta}, {Peoples},
  {Pier}, {Prieto}, {Prosapio}, {Rivetta}, {Schneider}, {Snedden}, \&
  {Wang}}]{Gunn2006}
{Gunn}, J.~E., {Siegmund}, W.~A., {Mannery}, E.~J., {et~al.} 2006, \aj, 131,
  2332, \dodoi{10.1086/500975}

\bibitem[{{Hern{\'a}ndez-Toledo} {et~al.}(2010){Hern{\'a}ndez-Toledo},
  {V{\'a}zquez-Mata}, {Mart{\'\i}nez-V{\'a}zquez}, {Choi}, \&
  {Park}}]{HernandezToledo2010}
{Hern{\'a}ndez-Toledo}, H.~M., {V{\'a}zquez-Mata}, J.~A.,
  {Mart{\'\i}nez-V{\'a}zquez}, L.~A., {Choi}, Y.-Y., \& {Park}, C. 2010, \aj,
  139, 2525, \dodoi{10.1088/0004-6256/139/6/2525}

\bibitem[{{Jacobs} {et~al.}(2019){Jacobs}, {Collett}, {Glazebrook}, {McCarthy},
  {Qin}, {Abbott}, {Abdalla}, {Annis}, {Avila}, {Bechtol}, {Bertin}, {Brooks},
  {Buckley-Geer}, {Burke}, {Carnero Rosell}, {Carrasco Kind}, {Carretero}, {da
  Costa}, {Davis}, {De Vicente}, {Desai}, {Diehl}, {Doel}, {Eifler},
  {Flaugher}, {Frieman}, {Garc{\'\i}a-Bellido}, {Gaztanaga}, {Gerdes},
  {Goldstein}, {Gruen}, {Gruendl}, {Gschwend}, {Gutierrez}, {Hartley},
  {Hollowood}, {Honscheid}, {Hoyle}, {James}, {Kuehn}, {Kuropatkin}, {Lahav},
  {Li}, {Lima}, {Lin}, {Maia}, {Martini}, {Miller}, {Miquel}, {Nord}, {Plazas},
  {Sanchez}, {Scarpine}, {Schubnell}, {Serrano}, {Sevilla-Noarbe}, {Smith},
  {Soares-Santos}, {Sobreira}, {Suchyta}, {Swanson}, {Tarle}, {Vikram},
  {Walker}, {Zhang}, {Zuntz}, \& {DES Collaboration}}]{Jacobs19}
{Jacobs}, C., {Collett}, T., {Glazebrook}, K., {et~al.} 2019, \mnras, 484,
  5330, \dodoi{10.1093/mnras/stz272}

\bibitem[{{Johnson} {et~al.}(2021){Johnson}, {Leja}, {Conroy}, \&
  {Speagle}}]{Johnson2021}
{Johnson}, B.~D., {Leja}, J., {Conroy}, C., \& {Speagle}, J.~S. 2021, \apjs,
  254, 22, \dodoi{10.3847/1538-4365/abef67}

\bibitem[{{Kauffmann} {et~al.}(2003){Kauffmann}, {Heckman}, {White}, {Charlot},
  {Tremonti}, {Brinchmann}, {Bruzual}, {Peng}, {Seibert}, {Bernardi},
  {Blanton}, {Brinkmann}, {Castander}, {Cs{\'a}bai}, {Fukugita}, {Ivezic},
  {Munn}, {Nichol}, {Padmanabhan}, {Thakar}, {Weinberg}, \&
  {York}}]{Kauffmann03}
{Kauffmann}, G., {Heckman}, T.~M., {White}, S. D.~M., {et~al.} 2003, \mnras,
  341, 33, \dodoi{10.1046/j.1365-8711.2003.06291.x}

\bibitem[{Kauffmann {et~al.}(2003)Kauffmann, Heckman, White, Charlot, Tremonti,
  Brinchmann, Bruzual, Peng, Seibert, Bernardi, Blanton, Brinkmann, Castander,
  Cs{\'{a}}bai, Fukugita, Ivezic, Munn, Nichol, Padmanabhan, Thakar, Weinberg,
  \& York}]{Kauffmann2003a}
Kauffmann, G., Heckman, T.~M., White, S. D.~M., {et~al.} 2003, Monthly Notices
  of the Royal Astronomical Society, 341, 33,
  \dodoi{10.1046/j.1365-8711.2003.06291.x}

\bibitem[{{Kewley} \& {Ellison}(2008)}]{Kewley08}
{Kewley}, L.~J., \& {Ellison}, S.~L. 2008, \apj, 681, 1183,
  \dodoi{10.1086/587500}

\bibitem[{Kroupa(2001)}]{Kroupa2001}
Kroupa, P. 2001, Monthly Notices of the Royal Astronomical Society, 322, 231,
  \dodoi{10.1046/j.1365-8711.2001.04022.x}

\bibitem[{Law {et~al.}(2016)Law, Cherinka, Yan, Andrews, Bershady, Bizyaev,
  Blanc, Blanton, Bolton, Brownstein, Bundy, Chen, Drory, D'Souza, Fu, Jones,
  Kauffmann, MacDonald, Masters, Newman, Parejko, S{\'{a}}nchez-Gallego,
  S{\'{a}}nchez, Schlegel, Thomas, Wake, Weijmans, Westfall, \&
  Zhang}]{Law2016}
Law, D.~R., Cherinka, B., Yan, R., {et~al.} 2016, The Astronomical Journal,
  152, 83, \dodoi{10.3847/0004-6256/152/4/83}

\bibitem[{Pietrinferni {et~al.}(2004)Pietrinferni, Cassisi, Salaris, \&
  Castelli}]{Pietrinferni2004}
Pietrinferni, A., Cassisi, S., Salaris, M., \& Castelli, F. 2004, The
  Astrophysical Journal, 612, 168, \dodoi{10.1086/422498}

\bibitem[{{Portillo} {et~al.}(2020){Portillo}, {Parejko}, {Vergara}, \&
  {Connolly}}]{Portillo}
{Portillo}, S. K.~N., {Parejko}, J.~K., {Vergara}, J.~R., \& {Connolly}, A.~J.
  2020, \aj, 160, 45, \dodoi{10.3847/1538-3881/ab9644}

\bibitem[{{Pourrahmani} {et~al.}(2018){Pourrahmani}, {Nayyeri}, \&
  {Cooray}}]{Pourrahmani18}
{Pourrahmani}, M., {Nayyeri}, H., \& {Cooray}, A. 2018, \apj, 856, 68,
  \dodoi{10.3847/1538-4357/aaae6a}

\bibitem[{{Rahmani} {et~al.}(2018){Rahmani}, {Teimoorinia}, \&
  {Barmby}}]{rahmani18}
{Rahmani}, S., {Teimoorinia}, H., \& {Barmby}, P. 2018, \mnras, 478, 4416,
  \dodoi{10.1093/mnras/sty1291}

\bibitem[{Rubin(1981)}]{Rubin81}
Rubin, D.~B. 1981, The Annals of Statistics, 9, 130 ,
  \dodoi{10.1214/aos/1176345338}

\bibitem[{{Salim} {et~al.}(2005){Salim}, {Charlot}, {Rich}, {Kauffmann},
  {Heckman}, {Barlow}, {Bianchi}, {Byun}, {Donas}, {Forster}, {Friedman},
  {Jelinsky}, {Lee}, {Madore}, {Malina}, {Martin}, {Milliard}, {Morrissey},
  {Neff}, {Schiminovich}, {Seibert}, {Siegmund}, {Small}, {Szalay}, {Welsh}, \&
  {Wyder}}]{Salim05}
{Salim}, S., {Charlot}, S., {Rich}, R.~M., {et~al.} 2005, \apjl, 619, L39,
  \dodoi{10.1086/424800}

\bibitem[{Sanchez {et~al.}(2012)Sanchez, Kennicutt, de~Paz, van~de Ven,
  V{\'{i}}lchez, Wisotzki, Walcher, Mast, Aguerri, Albiol-Perez,
  Alonso-Herrero, Alves, Bakos, Bartakova, Bland-Hawthorn, Boselli, Bomans,
  Castillo-Morales, Cortijo-Ferrero, de~Lorenzo-Caceres, del Olmo, Dettmar,
  D{\'{i}}az, Ellis, Falcon-Barroso, Flores, Gallazzi, Garc{\'{i}}a-Lorenzo,
  Delgado, Gruel, Haines, Hao, Husemann, Iglesias-P{\'{a}}ramo, Jahnke,
  Johnson, Jungwiert, Kalinova, Kehrig, Kupko, Lopez-Sanchez, Lyubenova,
  Marino, Marmol-Queralto, Marquez, Masegosa, Meidt, Mendez-Abreu,
  Monreal-Ibero, Montijo, Mourao, Palacios-Navarro, Papaderos, Pasquali,
  Peletier, Perez, Perez, Quirrenbach, Rela{\~{n}}o, Rosales-Ortega, Roth,
  Ruiz-Lara, Sanchez-Blazquez, Sengupta, Singh, Stanishev, Trager, Vazdekis,
  Viironen, Wild, Zibetti, \& Ziegler}]{Sanchez2012}
Sanchez, S.~F., Kennicutt, R.~C., de~Paz, A.~G., {et~al.} 2012, Astronomy {\&}
  Astrophysics, 538, A8, \dodoi{10.1051/0004-6361/201117353}

\bibitem[{{Sarmiento} {et~al.}(2021){Sarmiento}, {Huertas-Company}, \&
  {Knapen}}]{Sarmiento}
{Sarmiento}, R., {Huertas-Company}, M., \& {Knapen}, J.~H. 2021, in American
  Astronomical Society Meeting Abstracts, Vol.~53, American Astronomical
  Society Meeting Abstracts, 301.03

\bibitem[{Steidel {et~al.}(2014)Steidel, Rudie, Strom, Pettini, Reddy, Shapley,
  Trainor, Erb, Turner, Konidaris, Kulas, Mace, Matthews, \&
  McLean}]{Steidel2014}
Steidel, C.~C., Rudie, G.~C., Strom, A.~L., {et~al.} 2014, Astrophysical
  Journal, 795, 165, \dodoi{10.1088/0004-637X/795/2/165}

\bibitem[{{Teimoorinia}(2012)}]{teimoorinia12}
{Teimoorinia}, H. 2012, \aj, 144, 172, \dodoi{10.1088/0004-6256/144/6/172}

\bibitem[{{Teimoorinia} {et~al.}(2016){Teimoorinia}, {Bluck}, \&
  {Ellison}}]{Teimoorinia16}
{Teimoorinia}, H., {Bluck}, A. F.~L., \& {Ellison}, S.~L. 2016, \mnras, 457,
  2086, \dodoi{10.1093/mnras/stw036}

\bibitem[{{Teimoorinia} {et~al.}(2021{\natexlab{a}}){Teimoorinia},
  {Jalilkhany}, {Scudder}, {Jensen}, \& {Ellison}}]{teimoorinia21a}
{Teimoorinia}, H., {Jalilkhany}, M., {Scudder}, J.~M., {Jensen}, J., \&
  {Ellison}, S.~L. 2021{\natexlab{a}}, \mnras, 503, 1082,
  \dodoi{10.1093/mnras/stab466}

\bibitem[{{Teimoorinia} {et~al.}(2021{\natexlab{b}}){Teimoorinia},
  {Shishehchi}, {Tazwar}, {Lin}, {Archinuk}, {Gwyn}, \&
  {Kavelaars}}]{Teimoorinia21b}
{Teimoorinia}, H., {Shishehchi}, S., {Tazwar}, A., {et~al.} 2021{\natexlab{b}},
  \aj, 161, 227, \dodoi{10.3847/1538-3881/abea7e}

\bibitem[{{Teimoorinia} {et~al.}(2020){Teimoorinia}, {Toyonaga}, {Fabbro}, \&
  {Bottrell}}]{Teimoorinia20b}
{Teimoorinia}, H., {Toyonaga}, R.~D., {Fabbro}, S., \& {Bottrell}, C. 2020,
  \pasp, 132, 044501, \dodoi{10.1088/1538-3873/ab747b}

\bibitem[{{Tremonti} {et~al.}(2004){Tremonti}, {Heckman}, {Kauffmann},
  {Brinchmann}, {Charlot}, {White}, {Seibert}, {Peng}, {Schlegel}, {Uomoto},
  {Fukugita}, \& {Brinkmann}}]{Tremonti04}
{Tremonti}, C.~A., {Heckman}, T.~M., {Kauffmann}, G., {et~al.} 2004, \apj, 613,
  898, \dodoi{10.1086/423264}

\bibitem[{Vazdekis {et~al.}(2016)Vazdekis, Koleva, Ricciardelli, R{\"{o}}ck, \&
  Falc{\'{o}}n-Barroso}]{Vazdekis2016}
Vazdekis, A., Koleva, M., Ricciardelli, E., R{\"{o}}ck, B., \&
  Falc{\'{o}}n-Barroso, J. 2016, Monthly Notices of the Royal Astronomical
  Society, 463, 3409, \dodoi{10.1093/mnras/stw2231}

\bibitem[{Woo \& Ellison(2019)}]{Woo2019}
Woo, J., \& Ellison, S.~L. 2019, Monthly Notices of the Royal Astronomical
  Society, 487, 1927, \dodoi{10.1093/mnras/stz1377}

\bibitem[{{York} {et~al.}(2000){York}, {Adelman}, {Anderson}, {Anderson},
  {Annis}, {Bahcall}, {Bakken}, {Barkhouser}, {Bastian}, {Berman}, {Boroski},
  {Bracker}, {Briegel}, {Briggs}, {Brinkmann}, {Brunner}, {Burles}, {Carey},
  {Carr}, {Castander}, {Chen}, {Colestock}, {Connolly}, {Crocker}, {Csabai},
  {Czarapata}, {Davis}, {Doi}, {Dombeck}, {Eisenstein}, {Ellman}, {Elms},
  {Evans}, {Fan}, {Federwitz}, {Fiscelli}, {Friedman}, {Frieman}, {Fukugita},
  {Gillespie}, {Gunn}, {Gurbani}, {de Haas}, {Haldeman}, {Harris}, {Hayes},
  {Heckman}, {Hennessy}, {Hindsley}, {Holm}, {Holmgren}, {Huang}, {Hull},
  {Husby}, {Ichikawa}, {Ichikawa}, {Ivezi{\'c}}, {Kent}, {Kim}, {Kinney},
  {Klaene}, {Kleinman}, {Kleinman}, {Knapp}, {Korienek}, {Kron}, {Kunszt},
  {Lamb}, {Lee}, {Leger}, {Limmongkol}, {Lindenmeyer}, {Long}, {Loomis},
  {Loveday}, {Lucinio}, {Lupton}, {MacKinnon}, {Mannery}, {Mantsch}, {Margon},
  {McGehee}, {McKay}, {Meiksin}, {Merelli}, {Monet}, {Munn}, {Narayanan},
  {Nash}, {Neilsen}, {Neswold}, {Newberg}, {Nichol}, {Nicinski}, {Nonino},
  {Okada}, {Okamura}, {Ostriker}, {Owen}, {Pauls}, {Peoples}, {Peterson},
  {Petravick}, {Pier}, {Pope}, {Pordes}, {Prosapio}, {Rechenmacher}, {Quinn},
  {Richards}, {Richmond}, {Rivetta}, {Rockosi}, {Ruthmansdorfer}, {Sandford},
  {Schlegel}, {Schneider}, {Sekiguchi}, {Sergey}, {Shimasaku}, {Siegmund},
  {Smee}, {Smith}, {Snedden}, {Stone}, {Stoughton}, {Strauss}, {Stubbs},
  {SubbaRao}, {Szalay}, {Szapudi}, {Szokoly}, {Thakar}, {Tremonti}, {Tucker},
  {Uomoto}, {Vanden Berk}, {Vogeley}, {Waddell}, {Wang}, {Watanabe},
  {Weinberg}, {Yanny}, {Yasuda}, \& {SDSS Collaboration}}]{York2000}
{York}, D.~G., {Adelman}, J., {Anderson}, John~E., J., {et~al.} 2000, \aj, 120,
  1579, \dodoi{10.1086/301513}

\end{thebibliography}
\bibliographystyle{aasjournal}

\end{document}